\UseRawInputEncoding

\documentclass[conference]{IEEEtran}
\IEEEoverridecommandlockouts
\usepackage{cite}
\usepackage{amsmath,amsfonts,amsthm,bm,amssymb} %

\usepackage{algpseudocode,algorithm}
\usepackage{graphics}
\usepackage{epsfig}
\usepackage{footnote}
\usepackage{textcomp}
\PassOptionsToPackage{no-math}{fontspec}
\usepackage{xcolor}
\usepackage{tikz}
\usetikzlibrary{
    positioning,    % For 'right=of', 'above=of' node placement
    matrix,         % For multi-part nodes (matrix of nodes)
    arrows.meta,    % Modern arrow tips (Stealth, Latex etc.) - REPLACES 'arrows'
    shapes,         % Basic shapes
    chains,         % For chaining nodes automatically (optional)
    calc            % For coordinate calculations
}
\usepackage{makecell}
\usepackage{bm}
\usepackage{verbatim}
\usepackage{enumitem}
\def\BibTeX{{\rm B\kern-.05em{\sc i\kern-.025em b}\kern-.08em
		T\kern-.1667em\lower.7ex\hbox{E}\kern-.125emX}}
	
\usepackage{verbatim}
\usepackage{pgfplots}
\usepackage{hyperref}
\pgfplotsset{width=10cm,compat=1.9}
\pgfplotsset{
	every axis legend/.style={
		cells={anchor=center},
		inner xsep=3pt,inner ysep=2pt,
		nodes={inner sep=2pt,text depth=0.1em},
		anchor=north east,
		shape=rectangle,
		fill=white,draw=green,
		font=\footnotesize%\tiny
	},
}
\usepackage{standalone}
\usepackage{fancyhdr}
\usepackage{svg}
\usepackage{multirow}
\usepackage{subcaption}
\usepackage{graphicx}
\usepackage{enumitem}
\normalsize
\usepackage{tikz}
\usetikzlibrary{matrix, positioning, arrows.meta}
\usepackage{dsfont}
\newcommand{\uc}[1]{\textsc{\small #1}}
\usepackage{longtable}
\usepackage{booktabs}
\usepackage{tabularx}
\begin{document}
\title{Neural-Model-Augmented Hybrid NMS-OSD Decoders for Near-ML in Short Block Codes\\
%	\footnotesize 
%	\thanks{applicable funding agency here. If none, delete this.}
}
\begin{comment}
\author{\IEEEauthorblockN{Guangwen Li}
	\IEEEauthorblockA{\textit{College of Information \& Electronics} \\
		\textit{Shandong Technology and Business University}\\
		Yantai, China \\
		lgw.frank@sdtbu.edu.cn}
	\and
	\IEEEauthorblockN{Xiao Yu}
	\IEEEauthorblockA{\textit{Department of Physical Sports} \\
		\textit{Binzhou Medical University}\\
		Yantai, China \\
		yuxiao@bzmu.edu.cn}
}
\end{comment}

\author{Guangwen Li, Xiao Yu

\thanks{G.Li is with School of Information and Electronic Engineering, Shandong Technology and Business University, Yantai, China e-mail: lgw.frank@sdtbu.edu.cn}% <-this % stops a space
\thanks{X.Yu is with Teaching Department of Humanities and Social Sciences, Binzhou Medical University, Yantai, China e-mail: yuxiao@bzmc.edu.cn}% <-this % 
}

%\markboth{IEEE Transactions on Communications}%
%{Submitted paper}
\maketitle
\begin{abstract}
This paper presents a hybrid decoding architecture that serially couples a normalized min-sum (NMS) decoder with reinforced ordered statistics decoding (OSD) to achieve near-maximum likelihood (ML) performance for short linear block codes, including LDPC, BCH, and RS codes. The framework introduces several key innovations. A decoding information aggregation model based on a convolutional neural network refines bit-reliability estimates for OSD using the soft-output trajectory of the NMS decoder. An adaptive decoding path for OSD is initialized by the  arranged list of the most a priori likely tests algorithm and dynamically updated with empirical data. A sliding-window assisted model enables early termination of test error pattern (TEP) traversal, reducing complexity with minimal performance loss. For short high-rate codes, an undetected error detector identifies erroneous NMS outputs that satisfy parity checks, ensuring they are forwarded to OSD for correction. Extensive simulations on LDPC, BCH, and RS codes demonstrate that the proposed hybrid decoder achieves a competitive trade-off: near-ML frame error rate performance while maintaining advantages in throughput, latency, and complexity over state-of-the-art alternatives. Complexity analysis shows that the average number of OSD TEPs is drastically reduced, and the architecture remains highly parallelizable. An optimization framework is also formulated to balance performance and complexity via parameter tuning.
\end{abstract}


%\IEEEpeerreviewmaketitle
\section{Introduction}
\label{intro_sec}

Channel coding is fundamental to ensuring reliable information transmission in modern communication systems. Among linear block codes, low-density parity-check (LDPC) codes, first introduced by Gallager~\cite{gallager62}, are well known for approaching the Shannon limit asymptotically under belief propagation (BP) decoding~\cite{mackay96}. In complexity-sensitive applications, BP is often approximated by min-sum (MS) variants~\cite{zhao05,jiang06}. However, unlike maximum likelihood (ML) decoding, BP-type algorithms are suboptimal for finite-length LDPC codes due to short cycles in their Tanner graphs, and this performance gap can hardly be substantially bridged for the popular normalized min-sum (NMS)~\cite{chen2005reduced,ullah2011improved,ullah2011two} across the entire signal-to-noise ratio (SNR) range of interest. Moreover, extending these approaches to classical block codes with high-density parity-check (HDPC) matrices--such as Bose-Chaudhuri-Hocquenghem (BCH) and Reed-Solomon (RS) codes, which are widely employed in storage and deep-space communications for their excellent minimum-distance properties--remains challenging. Their rigid algebraic structure introduces numerous short cycles that violate the message-independence assumption underlying BP, resulting in significant performance degradation. Consequently, for both LDPC and HDPC codes, there is strong interest in developing decoding schemes that narrow the gap to ML while maintaining high throughput, low complexity, and low latency.

For classical block codes, exploiting algebraic properties to aid iterative decoding has been extensively studied. Jiang \emph{et al.}~\cite{jiang2004iterative} proposed stochastic shifting and adaptive damping for log-likelihood ratio (LLR) updates, and later reduced the parity-check matrix (PCM) based on bit reliability across iterations~\cite{jiang2006iterative}, preventing BP from stalling at pseudo-equilibria. The modified random redundant decoding algorithm~\cite{dimnik2009improved} employed fixed damping and multiple parallel decoders with permuted PCM, achieving lower complexity than the scheme in \cite{hehn2010multiple}, which relied on cyclic shifts of minimum-weight dual codewords. Halford \emph{et al.}~\cite{halford2006random} introduced random redundant decoding with a cycle-reduced PCM, while Ismail \emph{et al.}~\cite{ismail2015efficient} developed permuted BP by applying random automorphisms in each iteration. Santi \emph{et al.}~\cite{santi2018decoding} incorporated minimum-weight parity checks tailored to received sequences, improving BP for Reed-Muller (RM) codes at the cost of batch processing. Geiselhart \emph{et al.}~\cite{geiselhart2021automorphism} generalized ensemble decoding with diverse constituent decoders, achieving near-ML performance for RM codes. Deng \emph{et al.}~\cite{deng2020perturbed} proposed perturbed adaptive BP with advanced scheduling to accelerate convergence. However, PCM sparsification in these schemes often undermines BP's batch-processing capability and hampers hardware acceleration, such as on graphics processing units (GPUs).

Recognizing the central role of the PCM in BP performance, Lucas \emph{et al.}~\cite{lucas1998iterative} advocated constructing it from minimum-weight dual codewords to exploit sparsity. Yedidia \emph{et al.}~\cite{yedidia2002generating,shayegh2009low} proposed augmenting the PCM with auxiliary bits to reduce row weights and short cycles. Kou \emph{et al.}~\cite{kou01} showed that redundant PCM design can significantly improve BP for finite-length LDPC codes based on algebraic geometry, despite introducing additional short cycles. Inspired by this, similar strategies have been applied to HDPC codes~\cite{li2025effective}. For RS codes, embedding hard-decision decoding algorithms such as Berlekamp-Massey (BM) into each BP iteration has proven effective~\cite{shayegh2009low}. However, as in~\cite{deng2020perturbed}, these gains come at the expense of parallelizability.

Ordered statistics decoding (OSD)~\cite{Fossorier1995} provides a universal near-ML framework for linear block codes without relying on algebraic properties. Its drawback lies in the exponential growth of test error patterns (TEPs) with block length. Embedding OSD within BP~\cite{Fossorier1999,Fossorier2001} reduces TEPs per invocation but requires multiple OSD calls per codeword, creating a throughput bottleneck due to the mismatch between parallel BP and serial OSD. Recent work has focused on reducing OSD complexity while retaining frame error rate (FER) performance. Yue \emph{et al.}~\cite{Yue2021} introduced probabilistic early stopping for BCH codes and later mitigated Gaussian elimination overhead through preconditioning at high SNR region~\cite{Yue2022}. However, both approaches involve substantial probability computations tailored to each received sequence. Cavarec \emph{et al.}~\cite{cavarec2020learning} incorporated sufficient and necessary conditions for early stopping and skipping but faced long TEP traversals. Bossert \emph{et al.}~\cite{bossert2022hard} proposed information set decoding (ISD) for BCH codes under known noise variance, achieving competitive FER but limited throughput. Bailon \emph{et al.}~\cite{bailon2022concatenated} combined BM and OSD, yet both being serial, throughput bottlenecks remain.

Deep learning has transformed fields such as computer vision, natural language processing, and autonomous systems~\cite{lecun2015deep,bengio2021deep}. Recent advances in wireless communications leverage AI to address system imperfections, including robust direction-of-arrival estimation, AI-assisted dynamic frame structures for 5G networks, and robust resource allocation for physical layer security~\cite{labbaf2023robust,abedi2023ai,abedi2016robust}. In the context of channel coding, Ullah et al.~\cite{ullah2024otfs} highlight emerging trends, persistent challenges, and future research directions for OTFS-modulated massive MIMO systems employing 5G NR LDPC coding. Nachmani \emph{et al.}~\cite{nachmani16} pioneered neural belief propagation (NBP) by unrolling BP iterations into neural networks (NNs) with trainable message weights. Subsequent adaptations of MS variants improved decoding for classical codes~\cite{nachmani18,liang18,lugosch18,wang20,helmling19}. Lian \emph{et al.}~\cite{lian2019learned} showed that sharing parameters across SNRs in NBP reduces complexity without sacrificing performance. Other advances include teacher-student training~\cite{nachmani2022neural}, NN-pruned overcomplete parity-check matrices for short LDPC codes~\cite{buchberger20}, NBP with decimation~\cite{buchberger21}, and NBP ensembles tailored to absorbing sets~\cite{Rosseel2022}, achieving near-ML performance when concatenated with OSD. Simultaneously, Meenalakshmi \emph{et al.}~\cite{meenalakshmi2024deep} reviewed deep learning-based polar decoders for 5G and beyond, while Song \emph{et al.}~\cite{song2023one} proposed a self-supervised deep learning polar decoder enabling one-shot decoding without labeled data to reduce training overhead. Both highlight deep learning's potential to improve efficiency and adaptability in high-speed communication systems. However, the highly loopy structured Tanner graph underlying polar codes differs drastically from the randomly sparse graph for which BP was originally designed. The intervention of NBP only partially closes the performance gap due to the lack of mandatory code structure constraints in the definition of loss functions, limiting gains relative to the complexity overhead.

Improving bit reliability estimation for OSD has also been investigated. Jiang \emph{et al.}~\cite{jiang2007reliability} proposed weighted summation of posterior LLRs from NMS iterations, while Zhang \emph{et al.}~\cite{zhang2024efficient} refined reliability via additional BP-like iterations, albeit requiring accurate noise variance estimates. Babalola \emph{et al.}~\cite{babalola2019generalized} introduced generalized parity-check transformations for syndrome-based updates, but the resulting FER performance and parallelizability were limited. The success of ISD~\cite{bossert2022hard} also largely stems from its renewed bit reliability metrics derived from code-structure properties.

In our prior work~\cite{li2024boosting}, a hybrid NMS-OSD architecture was proposed for short LDPC codes, supported by two lightweight NNs: a decoding information aggregation (DIA) model to enhance bit reliability and a sliding-window assisted (SWA) model to timely terminate TEP traversal based on empirical orderings rather than Hamming weight. Subsequently, in~\cite{li2025effective}, we validated that the enhanced NMS (ENMS) can effectively improve the FER performance of BCH codes through deliberate redundancy, adaptation, and automorphism-based enhancements applied to the PCM and inputs.

\begin{figure}[t]
	\centering
	\includegraphics[width=0.42\textwidth]{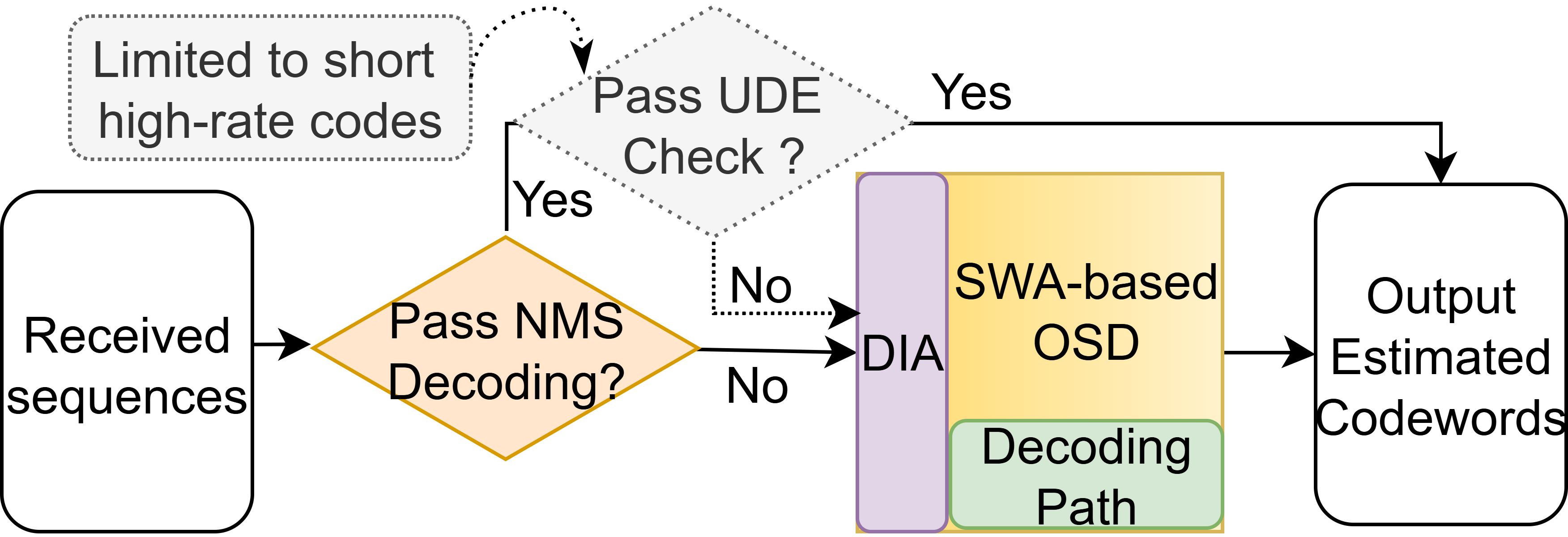}
\caption{Diagram of the proposed hybrid NMS-OSD architecture. 'NMS' refers to the standard NMS for LDPC codes or the ENMS for BCH/RS codes, depending on the code class.}
	\label{fig:general_nms_osd}
\end{figure}

In this paper, as illustrated in Fig.~\ref{fig:general_nms_osd}, we optimize and extend the existing NMS-OSD architecture to HDPC codes such as BCH and RS codes, alongside component improvements applicable to both LDPC and HDPC codes. The main contributions of this work are as follows:
\begin{itemize}
	\item The DIA and SWA models are adapted and validated for OSD decoding of BCH and RS codes.
	\item To address undetected decoding errors (UDEs) in NMS/ENMS for short high-rate codes, an optional UDE detection NN model is incorporated into the decoding architecture as shown in Fig.~\ref{fig:general_nms_osd}.
	\item The choice of hyperparameters affecting the FER performance of ENMS decoding of BCH and RS codes is elaborated.
	\item The optimized decoding path of OSD, consisting of blocks of TEPs of uniform size, outperforms its counterparts by requiring the fewest number of TEPs.
	\item Extensive validation of the hybrid NMS-OSD architecture across multiple code classes and lengths demonstrates favorable trade-offs among FER, throughput, complexity, and latency.
\end{itemize}

The rest of this paper is organized as follows. Section~\ref{preliminary} reviews the fundamentals of NMS, OSD, and Arranged List of the Most a Priori Likely Tests (ALMLT)~\cite{kabat2007new}, along with our prior work. Section~\ref{motivation} presents the methodological motivations. Section~\ref{simulations} provides experimental results and complexity analysis. Finally, Section~\ref{conclusions} concludes the paper and outlines directions for future research.
\section{Preliminaries}
\label{preliminary}

For convenience, the main symbols used throughout this paper are summarized in Table~\ref{tab:symbols}.
\begin{table}[htbp]
	\centering
	\caption{Summary of Symbols and Notations}
	\label{tab:symbols}
	\begin{tabular}{@{} l p{0.70\columnwidth} @{}}
		\toprule
		Symbol & Description \\
		\midrule
		$\mathbf{m}$ & Binary message vector of length $K$ \\
		$m_i$ & $i$-th bit of the binary message vector \\
		$\mathbf{c}$ & Codeword of length $N$ \\
		$c_i$ & $i$-th bit of the codeword \\
		$\mathbf{G}$ & Generator matrix of the code \\
		$\mathrm{GF}(2)$ & Galois field of two elements \\
		$K$ & Message length (in bits) \\
		$N$ & Codeword length (in bits) \\
		$s_i$ & BPSK-modulated symbol corresponding to $c_i$, i.e., $s_i = 1 - 2c_i$ \\
		$\mathbf{y}$ & Received sequence of length $N$ \\
		$y_i$ & Soft information of the $i$-th received symbol \\
		$n_i$ & Additive white Gaussian noise sample \\
		$\sigma^2$ & Noise variance \\
		$R$ & Code rate, $R = K/N$ \\
		$sp(\cdot)$ & Softplus function, $sp(x) = \log(1 + e^x)$ \\
		$\mathrm{sgn}(\cdot)$ & Sign function, returning $+1$ for $x \geq 0$ and $-1$ otherwise \\
		$l_{v_i}$ & Log-likelihood ratio for the $i$-th bit \\
		$\mathbf{1}(\cdot)$ & Indicator function \\
		$\mathbf{H}_{M \times N}$ & Parity-check matrix with $M$ rows and $N$ columns \\
		$M$ & Number of parity-check equations \\
		$I_m$ & Maximum number of NMS iterations \\
		$\alpha$ & Learned normalization parameter in NMS \\
		$\hat{\mathbf{c}}^{(t)}$ & Tentative hard decision at iteration $t$ \\
		$\hat{c}_i$ & Hard decision for the $i$-th bit \\
		$\pi_1$ & Permutation for bit reliability ordering in OSD \\
		$\pi_2$ & Permutation for Gaussian elimination in OSD \\
		$\mathbf{y}^{(2)}$ & Received sequence after permutations \\
		$p$ & Order of the OSD algorithm \\
		$\boldsymbol{\lambda}$ & Mean reliability vector for the ALMLT algorithm~\cite{kabat2007new} \\
		$\lambda_i$ & Mean reliability for the $i$-th bit position \\
		$l_t$ & Length of the ALMLT-based test error pattern list (decoding path) \\
		$I_s$ & Input dilation factor in ENMS for BCH codes \\
		$|s_n|$ & Number of cyclic shifts applied \\
		$|s_p|$ & Number of automorphism types applied \\
		$I_r$ & Redundancy factor (ratio of rows in $\mathbf{H}_o$ to original $\mathbf{H}$) \\
		$\mathbf{H}_o$ & Optimized parity-check matrix \\
		$C_r$ & Complexity ratio relative to a baseline NMS \\
		$\mathbf{d}$ & Discrepancy distance sequence for UDE detection \\
		$d_i$ & Weighted discrepancy between hard decisions at iteration $i$ and the final iteration \\
		$m_g$ & Margin for UDE detection, $m_g = p_a - p_r$ \\
		$p_a$ & Probability of acceptance from the UDE detector \\
		$p_r$ & Probability of rejection from the UDE detector \\
		$s_m$ & Soft margin for SWA-based OSD (difference in output probabilities) \\
		$d_m$ & Minimum distance of the code \\
		$\beta$ & Relaxing factor to decide the TEPs buffer size for decoding path \\
		$I_{at}$ & Average number of test error patterns traversed in OSD \\
		$b_s$ & Block size for partitioning the decoding path in SWA-based OSD \\
		$w_t$ & Window size (number of blocks) in SWA-based OSD \\
		$g_m$ & Global minimum distance in SWA-based OSD \\
		$I_{an}$ & Average number of NMS iterations \\
		$C_{hb}$ & Average complexity of the hybrid decoder \\
		$C_{n}$ & Complexity of the NMS decoder  \\
		$C_{o}$ & Complexity of the OSD decoder \\
		$F_{hb}$ & FER of the NMS-OSD decoder \\
		$F_n$ & FER of the NMS decoder \\
		$F_o$ & FER of the OSD decoder dealing with NMS decoding failures \\
		$d_v$ & Average column weight of the parity-check matrix \\
		$q$ & Number of quantization bits \\
		\bottomrule
	\end{tabular}
\end{table}

Given a binary row message vector $\mathbf{m} = [m_1, m_2, \ldots, m_K]$, the encoder maps it to a codeword $\mathbf{c} = [c_1, c_2, \ldots, c_N]$ via $\mathbf{c} = \mathbf{m}\mathbf{G}$ over $\mathrm{GF}(2)$, where $K$ and $N$ denote the message and codeword lengths, respectively. The generator matrix $\mathbf{G}$ is assumed to be full-rank without loss of generality.

After binary phase-shift keying (BPSK) modulation maps each codeword bit $c_i$ to a symbol $s_i = 1 - 2c_i$, the received sequence is $\mathbf{y} = [y_1, y_2, \ldots, y_N]$, where $y_i = s_i + n_i$ and $n_i$ is additive white Gaussian noise with zero mean and variance $\sigma^2$. The corresponding SNR in decibels is given by:
\[
\mathrm{SNR} = -10 \cdot \log_{10}(2 R \sigma^2),
\]
where $R = K/N$ is the code rate.

The LLR for the $i$-th bit is defined as:
\begin{equation}
	l_{v_i} = \log \left( \frac{p(y_i \mid c_i = 0)}{p(y_i \mid c_i = 1)} \right) = \frac{2y_i}{\sigma^2}.
	\label{llr_definition}
\end{equation}
Thus the magnitude of $y_i$ serves as a reliability metric for the hard decision $\mathbf{1}(y_i < 0)$, where $\mathbf{1}(\cdot)$ is the indicator function.

\subsection{NMS Decoding of LDPC codes}

The bipartite Tanner graph of a code is defined by its PCM $\mathbf{H}_{M \times N}$, consisting of $N$ variable nodes and $M$ check nodes connected by edges corresponding to the nonzero entries of $\mathbf{H}$.

For LDPC codes, NMS decoding~\cite{chen2005reduced} significantly reduces the computational complexity of BP while maintaining competitive error-rate performance. It also exhibits scale invariance~\cite{lugosch18}, allowing fixed noise variance assumptions (e.g., $\sigma^2 = 2$). The flooding schedule is adopted to enable fully parallel computation.

Let $t \in \{1, 2, \ldots, I_m\}$ denote the iteration index, where $I_m$ is the maximum number of iterations. The message from variable node $v_i$ to check node $c_j$ at iteration $t$ is:
\begin{equation}
	x_{v_i \to c_j}^{(t)} = l_{v_i} + \sum_{p \in \mathcal{C}(i) \setminus j} x_{c_p \to v_i}^{(t - 1)},
	\label{eq_v2c}
\end{equation}
where $\mathcal{C}(i) \setminus j$ denotes the set of check nodes connected to $v_i$ excluding $c_j$.

The message from check node $c_j$ to variable node $v_i$ is:
\begin{equation}
	x_{c_j \to v_i}^{(t)} = sp(\alpha) \cdot \min_{q \in \mathcal{V}(j) \setminus i} \left| x_{v_q \to c_j}^{(t)} \right| \cdot \prod_{q \in \mathcal{V}(j) \setminus i} \mathrm{sgn}\left( x_{v_q \to c_j}^{(t)} \right),
	\label{eq_c2v}
\end{equation}
where $sp(\cdot)$ and $\mathrm{sgn}(\cdot)$ denote the softplus and sign functions, respectively, and $\mathcal{V}(j) \setminus i$ denotes the set of variable nodes connected to $c_j$ excluding $v_i$. All messages $x_{c_p \to v_i}^{(0)}$ are initialized to zero, and the parameter $\alpha$ is learned when NMS is unrolled as a neural network~\cite{nachmani18}.

The posterior LLR for the $i$-th bit at iteration $t$ is:
\begin{equation}
	x_{v_i}^{(t)} = l_{v_i} + \sum_{p \in \mathcal{C}(i)} x_{c_p \to v_i}^{(t - 1)}.
	\label{eq_bit_message}
\end{equation}
A tentative hard decision $\hat{\mathbf{c}}^{(t)} = [\hat{c}_1, \hat{c}_2, \ldots, \hat{c}_N]$ is made via $\hat{c}_i = \mathbf{1}(x_{v_i}^{(t)} < 0)$ and checked for early termination using the syndrome condition:
\begin{equation}
	\mathbf{H} (\hat{\mathbf{c}}^{(t)})^\top = \mathbf{0}.
	\label{eq_early_termination}
\end{equation}

\subsection{OSD for Linear Block Codes}
\label{convention_osd}

Given $\mathbf{y}$ and a full-rank PCM $\mathbf{H}$, ordered statistics decoding~\cite{Fossorier1995} generates a list of candidate codewords and selects the most likely one. Unlike BP-based decoders, OSD is not affected by cycles in the Tanner graph and is applicable to any linear block code.

The order-$p$ OSD algorithm proceeds as follows:

\begin{itemize}
	\item Bit reliability ordering: Sort the bit indices in ascending order of reliability $|y_i|$ via a permutation $\pi_1$. The $K$ most reliable bits form the most reliable basis (MRB), and the remaining $N - K$ bits form the least reliable basis (LRB).
	
	\item Gaussian elimination: Apply column swaps (permutation $\pi_2$) to reduce $\pi_1(\mathbf{H})$ to systematic form:
	\[
	\pi_2 \cdot \pi_1(\mathbf{H}) = \mathbf{H}^{(2)} = [\mathbf{I} \mid \mathbf{Q}_2].
	\]
	Apply the same permutations to $\mathbf{y}$ to obtain $\mathbf{y}^{(2)} = \pi_2 \cdot \pi_1(\mathbf{y})$, and update the MRB and LRB accordingly.
	
	\item List decoding: For each test error pattern $\mathbf{e}_j$ of Hamming weight $\leq p$, compute a candidate codeword:
	\[
	\overline{\mathbf{c}}_{j,2} = \mathbf{c}_m \oplus \mathbf{e}_j, \quad \overline{\mathbf{c}}_{j,1} = \overline{\mathbf{c}}_{j,2} \mathbf{Q}_2,
	\]
	where $\mathbf{c}_m$ is the hard-decision estimate of the MRB. The candidate codeword $\overline{\mathbf{c}}_j = [\overline{\mathbf{c}}_{j,1} \mid \overline{\mathbf{c}}_{j,2}]$ is checked for validity via $\mathbf{H}^{(2)} \overline{\mathbf{c}}_j^\top = \mathbf{0}$. The optimal candidate is selected using:
	\begin{equation}
		\label{argmin_dis}
		\overline{\mathbf{c}} = \mathop {\arg\min }\limits_{\overline{\mathbf{c}}_j} \sum\limits_{i = 1}^N {\mathbf{1}(\check{c}_i \ne {c}_i)\left| {{y}^{(2)}_i} \right|},
	\end{equation}
		\begin{equation}
		\label{osd_dis}
		d_{\overline{\mathbf{c}}} = \mathop {\min }\limits_{\overline{\mathbf{c}}_j} 
		 \sum\limits_{i = 1}^N {\mathbf{1}(\check{c}_i \ne {c}_i)\left| {{y}^{(2)}_i} \right|},
	\end{equation}
	where $\overline{\mathbf{c}}_j = [c_i]_1^N$ and $\check{c}_i$ is the hard decision of $y^{(2)}_i$.
	
	\item Inverse permutation: Apply $\pi_1^{-1} \circ \pi_2^{-1}$ to $\overline{\mathbf{c}}$ to obtain the final codeword estimate $\hat{\mathbf{c}}$.
\end{itemize}

\subsection{ALMLT Algorithm}

A major drawback of conventional OSD is its exponential complexity with respect to the decoding order $p$. The ALMLT algorithm~\cite{kabat2007new} alleviates this issue by prioritizing TEPs according to statistical reliability, and has been shown effective for long block codes. It proceeds as follows:

\begin{itemize}
	\item Compute a mean reliability vector $\bm{\lambda} = [\lambda_1, \lambda_2, \ldots, \lambda_N]$ under the AWGN assumption using ordered statistics, while ignoring the Gaussian elimination permutation $\pi_2$.
	\item For a given SNR, arrange TEPs in ascending order of their weight, where the weight is defined as the sum of $\lambda_i$ over the flipped positions within the MRB section. This list is precomputed and remains fixed during decoding. 
\end{itemize}

Empirical results in~\cite{kabat2007new} demonstrate that OSD guided by an ALMLT list of arbitrary length $l_t$ often achieves better FER performance than conventional order-$p$ OSD, which uses a trimmed list of the same number of TEPs selected solely by Hamming weight.

\subsection{Our Prior Work}

\subsubsection{DIA and SWA Models}

\begin{figure}[htbp]  
	\centering
	\includegraphics[width=0.49\textwidth]{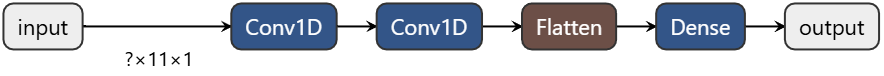}
	\caption{DIA model architecture, where '$? \times 11 \times 1$' denotes (batch size) $\times$ (1-D posterior LLR list of length $I_m+1=11$), with the additional entry arising from the raw input sequence.}
	\label{fig:model_dia}
\end{figure}

In the hybrid NMS-OSD architecture (Fig.~\ref{fig:general_nms_osd}), OSD is invoked only when NMS decoding fails. In such cases, the trajectories of posterior LLRs from NMS decoding are available to prepare enhanced reliability measurements for OSD decoding. To this end, as shown in Fig.~\ref{fig:model_dia}, the DIA model comprises two 1D convolutional layers with two and one filters of size 3 across iterations, respectively, followed by a flattening layer to accommodate possibly varying $I_m$ settings, and a dense layer to output renewed bit soft information.

\begin{figure}[htbp]  
	\centering
	\includegraphics[width=0.4\textwidth]{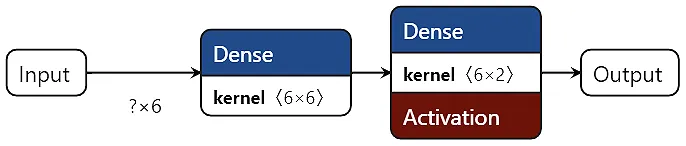}
	\caption{SWA model architecture, where '$? \times 6$' denotes (batch size) $\times$ (1-D input vectors of length $6$), with the six entries comprising a consecutive minimum-distance list of length $w_t = 5$ and the current window position.}
	\label{fig:model_swa}
\end{figure}

To manage the decoding complexity of OSD, a fixed long TEP list, named the decoding path, is first acquired by statistically counting the sum of hits of true error patterns within each TEP block, and then rearranging the blocks in descending order of the ratio of hits to block size. Then the SWA model~\cite{li2024boosting}, a two-layer neural network implementation, is called to determine when to early terminate the traversal along the decoding path. As shown in Fig.~\ref{fig:model_swa}, the softmax activation function of the second layer assesses the risk of early termination in the form of a probability difference.

\subsubsection{ENMS decoding for BCH Codes}

BCH codes remain competitive owing to their strong minimum distance properties. To enable high-throughput decoding, an ENMS decoder was proposed in~\cite{li2025effective}, which fully exploits the cyclic structure of BCH codes. A prerequisite for this approach is to transform the original PCM $\mathbf{H}$ into an optimized form $\mathbf{H}_o$, thereby improving sparsity and other Tanner graph properties such as a quasi-uniform column weight distribution beneficial for decoding.

The ENMS decoder is summarized in Algorithm~\ref{alg::enhanced_nms}. In Step~2, $|s_n|$ and $|s_p|$ denote the number of shifts and automorphism types applied to each sequence, effectively expanding the input by a dilation factor $I_s = |s_n| |s_p|$. In Step~10, to accommodate the merging operation, the original received vector $\mathbf{y}$ is discarded and reassigned with the posterior LLRs obtained after each NMS iteration, resulting in accelerated decoding convergence.

\begin{algorithm}[ht]
	\caption{ENMS Decoding of BCH Codes}
	\label{alg::enhanced_nms}
	\begin{algorithmic}[1]
		\Require Received $\mathbf{y}$, $\mathbf{H}_o$, fixing $\sigma^2 = 2$ so that $l_{v_i}=y_i$
		\Ensure Estimated codeword $\hat{\mathbf{c}}$
		\State \textbf{for} $t = 1$ to $I_m$
		\State \quad\textbf{Dilate:} Expand $\mathbf{y}$ into a block $B(\mathbf{y})$ of size $I_s \times N$ after applying all shifts and automorphisms.
		\State \quad\textbf{Parallel Message Passing:}
		\State \quad\quad Compute V2C messages for $B(\mathbf{y})$ via~\eqref{eq_v2c}, except removing the second term on the right-hand side.
		\State \quad\quad Compute C2V messages for $B(\mathbf{y})$ via~\eqref{eq_c2v}.
		\State \quad\textbf{Aggregate:} Reverse all permutations on the messages from~\eqref{eq_c2v} and sum the results for each variable node $v_i$.
		\State \quad\textbf{Fuse:} Compute the posterior LLR $x_{v_i}^{(t)}$ by substituting the aggregated sum into the second term of the right-hand side of~\eqref{eq_bit_message}.
		\State \quad Obtain hard decision $\hat{\mathbf{c}}^{(t)}$ where $\hat{c}_i = \mathbf{1}(x_{v_i}^{(t)} < 0)$.
		\State \quad\textbf{if} $\mathbf{H}_o (\hat{\mathbf{c}}^{(t)})^\top = \mathbf{0}$ \textbf{then break}
		\State \quad\textbf{else} $l_{v_i} \gets x_{v_i}^{(t)}$, $i \in [1,N]$. \Comment{Update!}
		\State \textbf{end for}
		\State \Return $\hat{\mathbf{c}}^{(t)}$
	\end{algorithmic}
\end{algorithm}
\section{Motivations}
\label{motivation}
Hereafter, for LDPC codes, NMS refers to standard NMS operating on the original PCM $\mathbf{H}$~\cite{chen2005reduced}, whereas for BCH and RS codes, the ENMS operating on the optimized $\mathbf{H}_o$ is assumed by default. This section first provides detailed training nuances for all models to ensure reproducibility. It then validates the NN-based UDE detector for short high-rate codes. Next, we discuss the impact of hyperparameter tuning on NMS decoding of BCH and RS codes. Finally,  after devising the decoding path update mechanism, we extend the DIA model across code classes, and validate the performance of the SWA-based OSD to traverse TEP blocks of uniform size along the decoding path.

\subsection{Training logistics of NN models}
There are two classes of NN models to be trained in supervised learning mode: NMS decoders and auxiliary models such as the UDE, DIA, and SWA. To align with the noise-agnostic property of NMS and OSD, the training dataset for all NN models is acquired at a specific SNR, then the trained models function across any SNR of interest.

The following setups are assumed by default, unless stated otherwise. For any model, a training dataset of size $10^4$ is prepared at the SNR that yields an FER of approximately 0.1 for the optimized NMS. The Adam optimizer~\cite{kingma2014} is applied to optimize the cross-entropy loss on training samples with a batch size of 100, an initial learning rate of 0.01 that decays by 0.95 every 100 steps, and a termination step of $10^4$. Notably, all trainings are executed on a standard computer with a 2.60 GHz Intel i7-6700HQ processor.

\subsubsection{NMS models}
For any LDPC code, the sole parameter $\alpha$ of NMS is slightly affected by the $I_m$ setup. Combined with the transmitted codewords as the ground-truth labels, randomly generated received sequences constitute the training dataset. The training is expected to minimize the loss function within 30 minutes after reaching the termination step of $10^3$. Similar training steps are applied for any BCH or RS code, except that the only trainable $\alpha$ of its NMS is closely coupled with the $(I_s,I_r,I_m)$ setups; any variation requires retraining the NMS-associated $\alpha$ in the new configuration.

\subsubsection{Training of various auxiliary models}
The DIA model \cite{li2024boosting}, shown in Fig.~\ref{fig:model_dia}, is a four-layer convolutional neural network (CNN)~\cite{chollet2015keras} employing ReLU activation for the first two Conv1D layers and linear activation for the final Dense layer. For LDPC codes, the first two Conv1D layers are configured with '2' and '1' filters, respectively, while for BCH and RS codes, '4' and '2' filters are used to enhance model capacity. The training data is derived from NMS decoding failures. Specifically, for each decoding failure, the posterior LLRs of every codeword bit across the iterative NMS decoding process are paired with labels corresponding to the transmitted codeword bits. This yields a total of $N$ trajectories, each of length $I_m+1$, where the ``+1'' accounts for the soft information of bits from the original input sequence. The head element of each trajectory is then rectified to be always positive by multiplying its sign across all trajectory elements, forming one training sample. Consequently, a training dataset of size $N \times 10^4$ can be readily obtained. The outputs of the trained DIA model are subsequently rectified by multiplying them with the corresponding signs. Under default training settings, the optimization process is expected to complete within 10 minutes.

The SWA model as shown in Fig.~\ref{fig:model_swa}, closely embedded in OSD decoding, operates on blocks of TEPs of size $b_s$ after sequentially and uniformly partitioning all TEPs of the decoding path. The preparation of its training dataset is more complex than that of the other models. After concatenating the DIA model outputs with the original channel sequence, each codeword bit has two reliability measurements. The former provides evidence for the $\pi_1$ permutation and hard decision of the MRB in OSD, while the latter is dedicated to any operations involving distance calculation such as \eqref{argmin_dis} or \eqref{osd_dis} (see \cite{li2024boosting} for more details of SWA-based OSD). Hence, given a decoding path comprising blocks $\bm{b}_i, i=1,2,\cdots,l_p$, applying Equation~\eqref{osd_dis} to the TEPs belonging to each $\bm{b}_i$ generates a list of $l_p$ minima. Sliding a window of width $w_t$ over this list with a default stride of one creates $m_s = l_p - w_t + 1$ training samples, using every $w_t$ minima per sample. For better decision accuracy, the sliding index of the current window is appended to the end of the list of $w_t$ elements sorted in ascending order to form one training sample. Thus, a total training sample size of $m_s \cdot 10^4$ is sufficient as the training dataset for the SWA model. If the true minimum distance corresponding to the transmitted codeword lies within the current window, the one-hot encoded label '0' is assigned to terminate further decoding. Otherwise, the one-hot encoded label '1' signals proceeding with decoding.

The SWA model incorporates a weighted cross-entropy loss to address class imbalance and penalize incorrect predictions of premature termination in \eqref{swa_ce_loss}:

\begin{equation}
	\label{swa_ce_loss}
	\ell_{\text{ce}} = \sum_{i=1}^{b_n} \omega_{1(2)} \beta_i \left( \sum_{j=0}^1 p(d_i = j) \cdot \log \frac{1}{p(\hat{d}_i = j)} \right)
\end{equation}

In this equation, as the class weight, $\omega_1$ or $\omega_2$ is assigned to the $i$-th sample depending on whether it is a positive or negative sample within a batch of training data of size $b_n$ sampled from the training dataset. The ratio of $\omega_1$ and $\omega_2$ should be chosen to rectify the imbalance in the number of positive and negative samples. The penalty factor $\beta_i$ is calculated as $\max(1, \gamma \cdot \mathbf{1}(d_i = 1, \hat{d}_i = 0))$, with $\gamma$ set to 10. This configuration imposes a significant penalty on predictions that incorrectly identify a '1' as '0', thereby minimizing adverse effects on the FER. Conversely, misclassifications where the prediction is '1' for a true label of '0' are not subject to this penalty, as they are deemed to have no impact on FER. We once again train the SWA model with the previously declared default training settings, and the training is expected to converge after reaching the termination step of $20,000$ within 20 minutes.

For the UDE model shown in Fig.~\ref{fig:ude_nn}, the NMS decoding trajectories are collected that satisfy all parity checks of $\mathbf{H}_o$ (or $\mathbf{H}$) at the SNR where the resulting FER is about 0.3. Then the discrepancy distance list $\mathbf{d}=[d_i]_{i=0}^{I_m-1}$ is obtained with $d_i$ defined in \eqref{ude_distance}:
\begin{equation}
	d_i = \sum_{j=1}^{N} \mathbf{1}\!\left(\hat{c}_j^{(i)} \neq \hat{c}_j^{(I_m)}\right) |y_j|,
	\label{ude_distance}
\end{equation}
where $\hat{c}_j^{(i)}$ denotes the hard decision of the $j$-th component of the posterior LLR at iteration $i$, and $|y_j|$ is the magnitude of the received symbol. Hence, the $d_i$ metric measures the soft discrepancy between the $i$-th and final $I_m$-th iterations. Intuitively, the discrepancy distance list exhibits a smooth and gradual converging trend for correct decodings (true positives) and an erratic trend for UDEs (false positives). Each discrepancy distance list is treated as a training sample and labeled as a positive case with a one-hot encoded label when the hard decision of the $I_m$-th NMS decoding matches the transmitted codeword. After applying the default training settings declared previously, the optimization is expected to complete within 10 minutes after reaching the termination training step.

\subsection{UDE Model}
For short high-rate codes, the UDEs of NMS decoding--where all parity-check constraints are satisfied but the decoding is incorrect--occur with non-negligible probability, especially in the low- to medium-SNR regime, due to the limited minimum distance. In a hybrid NMS-OSD architecture, such UDEs degrade the FER performance because escaping detection by NMS subsequently prevents the triggering of OSD. To address this issue, we propose a UDE detection NN model that decides whether to accept or reject an NMS decoding even when all parity-check constraints of $\mathbf{H}$/$\mathbf{H}_o$ are satisfied. That is, the UDE model arbitrates the NMS decoding based on the trend of the available distance trajectory during iterative NMS decoding.

\begin{figure}[htbp]
	\centering
        \resizebox{0.48\textwidth}{!}{  % Scales width to 48% of text width, height adjusts proportionally
    \begin{tikzpicture}[
    node distance=0.5cm,
    layer/.style={
        rectangle, draw=black!60, thick,
        minimum width=2cm, minimum height=1cm,
        align=center
    },
    multipart/.style args={#1/#2/#3}{
        rectangle, draw=black!60, very thick,
        minimum width=2cm, minimum height=1cm,
        path picture={
            \fill[#1] (path picture bounding box.north west) 
                rectangle ([yshift=-0.33cm]path picture bounding box.north east);
            \fill[#2] ([yshift=-0.33cm]path picture bounding box.north west) 
                rectangle ([yshift=-0.66cm]path picture bounding box.north east);
            \fill[#3] ([yshift=-0.66cm]path picture bounding box.north west) 
                rectangle (path picture bounding box.south east);
        }
    },
    arrow/.style={-Stealth, thick}
]

% Standard nodes
\node [layer, fill=blue!10] (input) {Input};
\node [layer, fill=green!10, right=of input] (conv) {Conv1D \\ kernel (2$\times$3$\times$1)};
\node [layer, fill=yellow!10, right=of conv] (flatten) {Flatten};
\node [layer, fill=green!10,  right=of flatten] (dense) {Dense \\ Softmax:($?\times2$) };
\node [layer, fill=purple!10, right=of dense] (output) {Output};

% Arrows
\draw [arrow] (input) -- (conv);
\draw [arrow] (conv) -- (flatten);
\draw [arrow] (flatten) -- (dense);
\draw [arrow] (dense) -- (output);
\end{tikzpicture}
}
\caption{UDE model architecture where '$2\times 3 \times 1$' denotes (\# of filters) $\times$ (kernel size) $\times$ (stride) and '$?\times 2$' means (batch size) $\times$ (output dimension).}
\label{fig:ude_nn}
\end{figure}

As shown in Fig.~\ref{fig:ude_nn}, the proposed three-layer UDE model adopts the softmax activation function to output a binary probability distribution: acceptance probability $p_a$ and rejection probability $p_r$. A predetermined margin $m_g = p_a - p_r$ serves as the decision threshold, and its variation can impact the balance between performance and complexity. Compared to direct thresholding on distance, the proposed model--trained at a specific SNR point--achieves robust performance across all SNRs of interest. We postpone the training of the UDE model to the subsection below for comparison with other models.

\begin{figure}[htbp]
	\centering
	\begin{subfigure}[b]{0.24\textwidth}
		\centering
		\includegraphics[width=\textwidth]{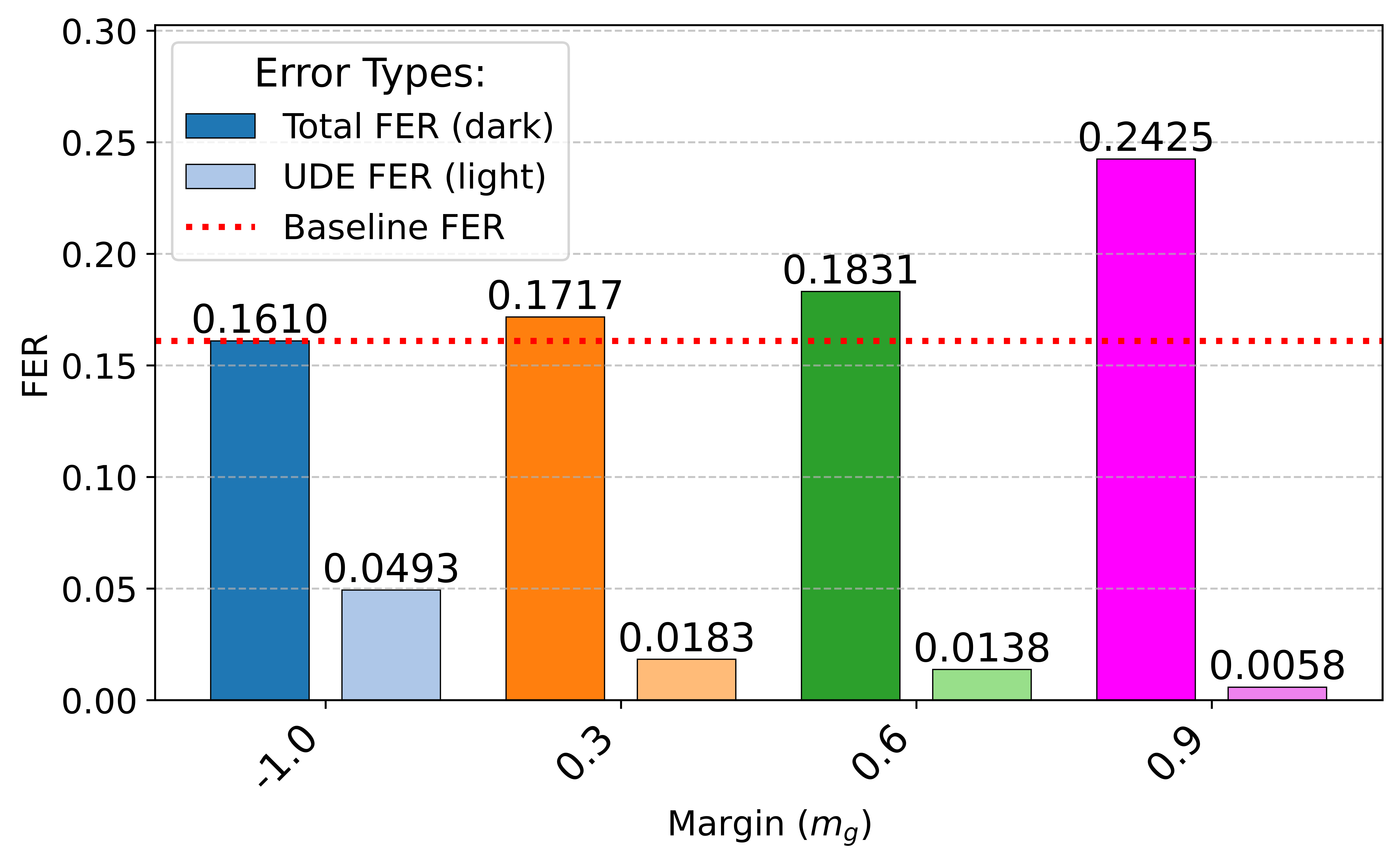}
		\caption{Training at ${\rm SNR}=2.6$~dB}
		\label{fig:ude_1a}
	\end{subfigure}
	\hfill
	\begin{subfigure}[b]{0.24\textwidth}
		\centering
		\includegraphics[width=\textwidth]{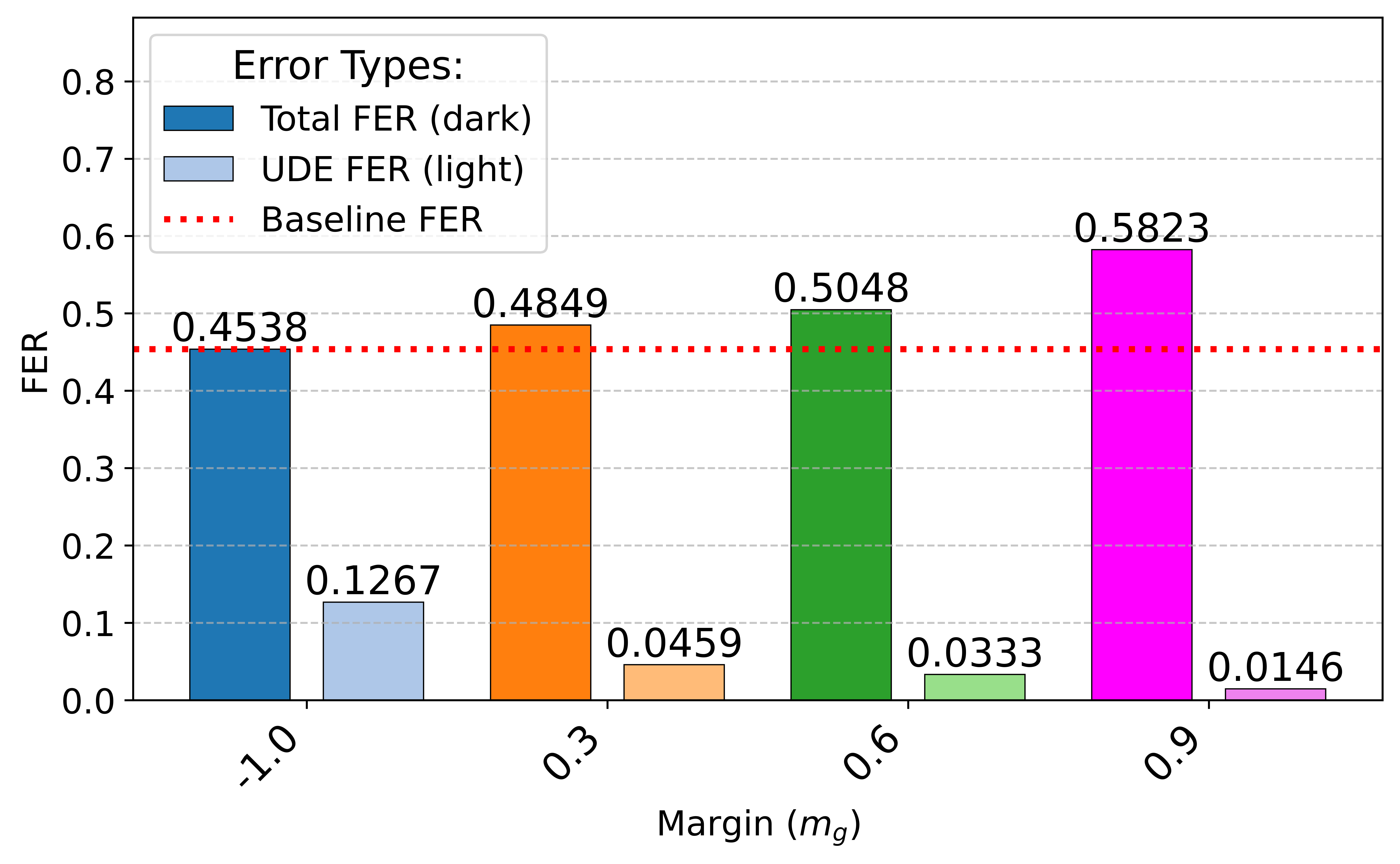}
		\caption{Testing at ${\rm SNR}=1.5$~dB}
		\label{fig:ude_1b}
	\end{subfigure}
	\caption{Detection performance and robustness of the UDE model on the BCH (63,45) code.}
	\label{fig:ude_2.6dB}
\end{figure}

For the BCH (63,45) code with $(I_s,I_r,I_m)=(9,2,4)$ at ${\rm SNR}=2.6$~dB, where $I_r$ measures the redundancy degree of $\mathbf{H}_o$, Fig.~\ref{fig:ude_1a} shows that when validating at the same SNR as training, without UDE detection (denoted by $m_g=-1$), UDEs contribute an FER of 0.049--nearly one-third of the total FER of 0.161. Increasing $m_g$ reduces the UDE fraction but introduces false negatives (i.e., correct decodings incorrectly rejected), which adds an excess FER relative to the $m_g=-1$ baseline. Importantly, Fig.~\ref{fig:ude_1b} demonstrates that the model remains effective when testing is performed at a far drifted SNR of $1.5$~dB.

\begin{table}[ht]
	\centering
	\caption{Detection performance of the UDE model on RS codes}
	\label{tab:RS_UDE}
	\resizebox{\columnwidth}{!}{%
		\begin{tabular}{|c|c|cccc|}
			\hline
			&
			&
			\multicolumn{4}{c|}{Margin $m_g$} \\ \cline{3-6} 
			\multirow{-2}{*}{Codes} &
			\multirow{-2}{*}{Metric} &
			\multicolumn{1}{c|}{$-1$} &
			\multicolumn{1}{c|}{$0.3$} &
			\multicolumn{1}{c|}{$0.6$} &
			$0.9$ \\ \hline
			&
			Total FER &
			\multicolumn{1}{c|}{{\color[HTML]{3531FF} 0.3280}} &
			\multicolumn{1}{c|}{{\color[HTML]{3531FF} 0.3297}} &
			\multicolumn{1}{c|}{{\color[HTML]{3531FF} 0.3332}} &
			{\color[HTML]{3531FF} 0.3586} \\ \cline{2-6} 
			\multirow{-2}{*}{\begin{tabular}[c]{@{}c@{}}RS (15,11) code\\ at SNR=2.6dB\end{tabular}} &
			UDE FER &
			\multicolumn{1}{c|}{{\color[HTML]{3531FF} 0.01765}} &
			\multicolumn{1}{c|}{{\color[HTML]{3531FF} 0.01152}} &
			\multicolumn{1}{c|}{{\color[HTML]{3531FF} 0.0097}} &
			{\color[HTML]{3531FF} 0.00578} \\ \hline
			&
			Total FER &
			\multicolumn{1}{c|}{{\color[HTML]{3531FF} 0.2631}} &
			\multicolumn{1}{c|}{{\color[HTML]{3531FF} 0.3331}} &
			\multicolumn{1}{c|}{{\color[HTML]{3531FF} 0.3888}} &
			{\color[HTML]{3531FF} 1.0000} \\ \cline{2-6} 
			\multirow{-2}{*}{\begin{tabular}[c]{@{}c@{}}RS (15,13) code\\ at SNR=3.0dB\end{tabular}} &
			UDE FER &
			\multicolumn{1}{c|}{{\color[HTML]{3531FF} 0.21948}} &
			\multicolumn{1}{c|}{{\color[HTML]{3531FF} 0.07913}} &
			\multicolumn{1}{c|}{{\color[HTML]{3531FF} 0.05188}} &
			{\color[HTML]{3531FF} 0.0000} \\ \hline
			&
			Total FER &
			\multicolumn{1}{c|}{{\color[HTML]{3531FF} 0.3323}} &
			\multicolumn{1}{c|}{{\color[HTML]{3531FF} 0.3897}} &
			\multicolumn{1}{c|}{{\color[HTML]{3531FF} 0.4461}} &
			{\color[HTML]{3531FF} 0.6683} \\ \cline{2-6} 
			\multirow{-2}{*}{\begin{tabular}[c]{@{}c@{}}RS (31,29) code\\ at SNR=4.0dB\end{tabular}} &
			UDE FER &
			\multicolumn{1}{c|}{{\color[HTML]{3531FF} 0.22099}} &
			\multicolumn{1}{c|}{{\color[HTML]{3531FF} 0.0632}} &
			\multicolumn{1}{c|}{{\color[HTML]{3531FF} 0.04103}} &
			{\color[HTML]{3531FF} 0.00707} \\ \hline
		\end{tabular}
	}
\end{table}

The UDE model was further evaluated on the binary images of RS (15,11) and RS (15,13) codes under the configuration $(I_s,I_r,I_m)=(5,2,4)$, and the RS (31,29) code under $(5,2,8)$. The testing results are shown in Table~\ref{tab:RS_UDE}. For the RS (15,11) code with known minimum distance $d_m=8$, the UDE FER of about 0.0177 is almost negligible compared to the total FER of 0.328 at $m_g=-1$, suggesting that the detection step using the UDE model may be skipped with minimal loss. In contrast, the UDE model is indispensable for discriminating UDE cases for the other two codes. For the RS (15,13) code with $d_m=4$, the FER performance of the UDE model is highly sensitive to the choice of $m_g$: the UDE FER drops from about 0.219 to 0.079 as $m_g$ increases from $-1$ to 0.3; at an inappropriate $m_g=0.9$, all NMS decoding results are rejected, forcing OSD to handle all decodings from scratch. Hence, $m_g=0.3$ is practical in this context. A similar trend is observed for the RS (31,29) code, which also has $d_m=4$.

Notably, for the RS (15,13) code with $m_g=0.3$, nearly half of the rejected cases--which contribute to a total FER of 0.33--will fail OSD post-processing, since the overall FER performance is bounded by the ML FER of 0.198 at ${\rm SNR}=3.0$~dB~\cite{helmling19}.

Beyond adjusting $m_g$ to balance the ratio of false positives and false negatives, the UDE model can be tuned toward conservative or aggressive detection by modifying class weights during training.
Overall, the UDE model provides a quantitative confidence measure for NMS or other iterative decodings and generalizes well across codes and various SNR points.
\subsection{Configurations for NMS of BCH and RS codes}
For BCH or RS codes, according to our implementation, the trainable variable $\alpha$ and the FER performance of NMS decoding depend critically on the setup of the tuple $(I_s,I_r,I_m)$. The input dilation factor $I_s = |s_n||s_p|$, where $|s_p|$ denotes the number of available automorphism types; thus $|s_p|=3$ and $1$ for BCH and RS codes, respectively, while $|s_n|$ denotes the number of cyclic shifts adopted, which is empirically chosen from 1 to $N/6$. The special value $I_s = -1$ indicates no cyclic shift. The redundancy factor $I_r$ quantifies the ratio of the number of rows in the optimized $\mathbf{H}_o$ relative to the original $\mathbf{H}$. Finally, the maximum number of iterations $I_m$ specifies the decoding depth of NMS.

\begin{figure}[htbp]
	\centering
	\begin{minipage}[t]{0.24\textwidth}
		\centering
		\includegraphics[width=\textwidth]{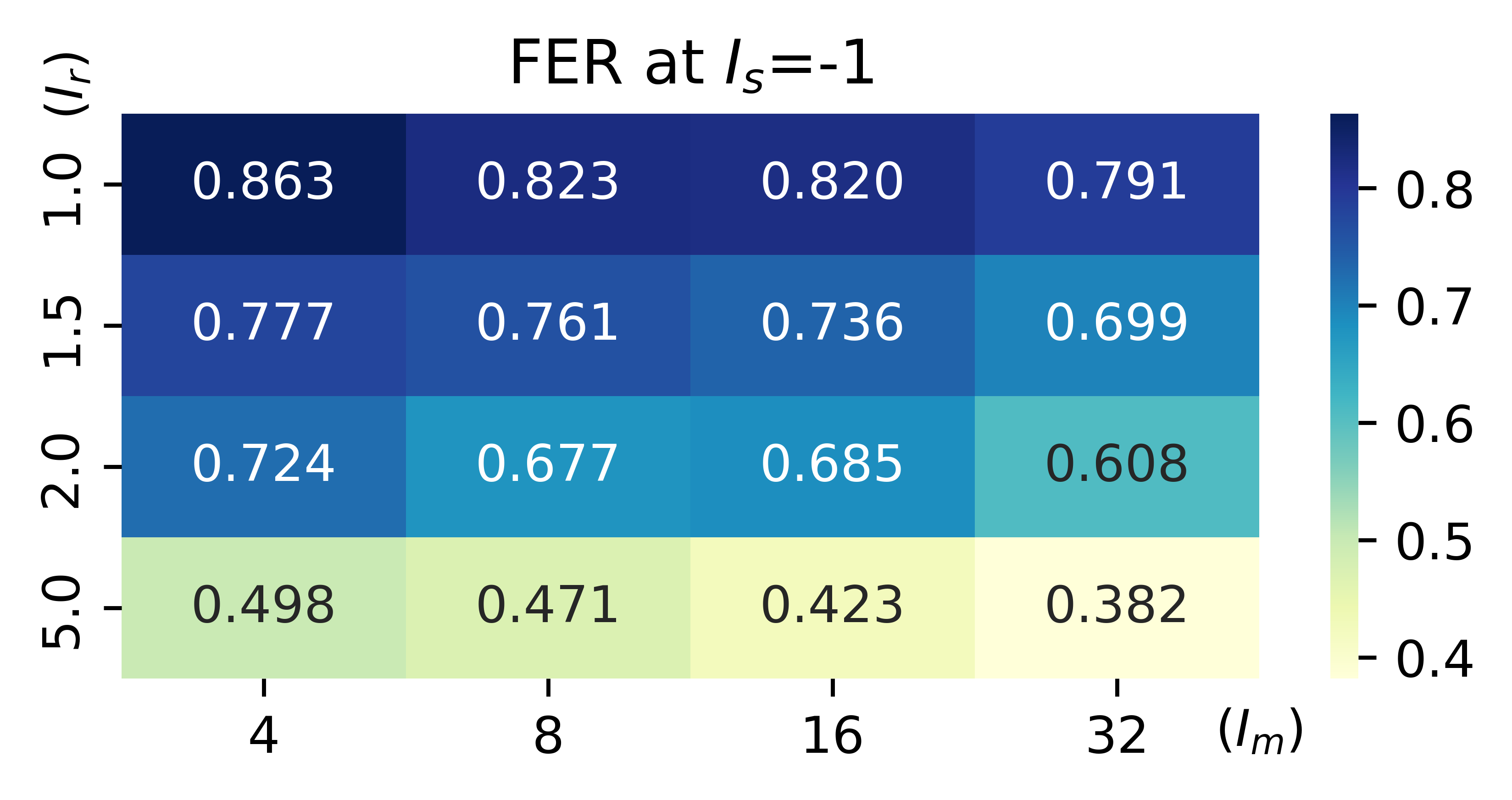}
		\vspace{0.2cm}
		\includegraphics[width=\textwidth]{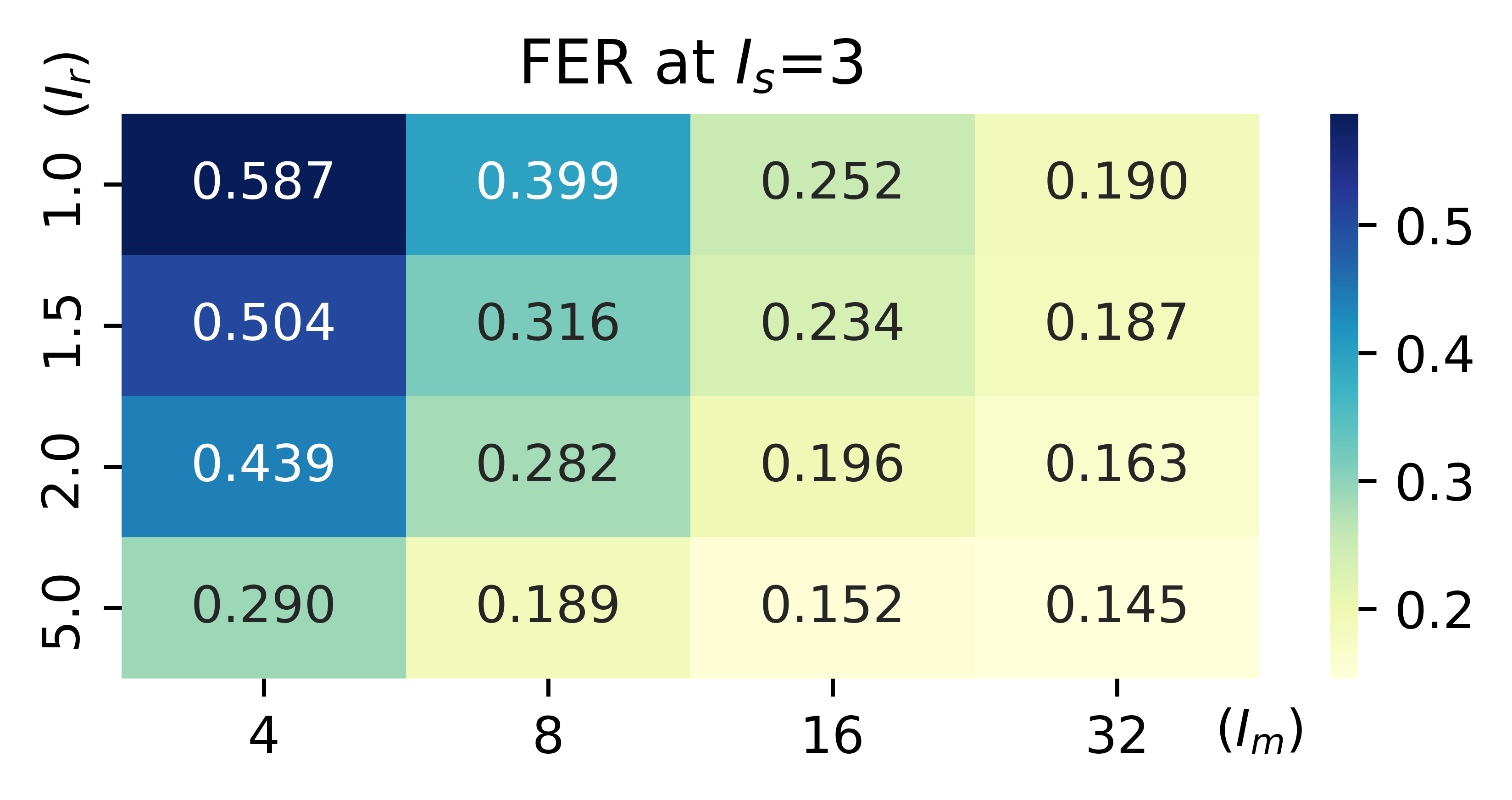}
		\vspace{0.2cm}
		\includegraphics[width=\textwidth]{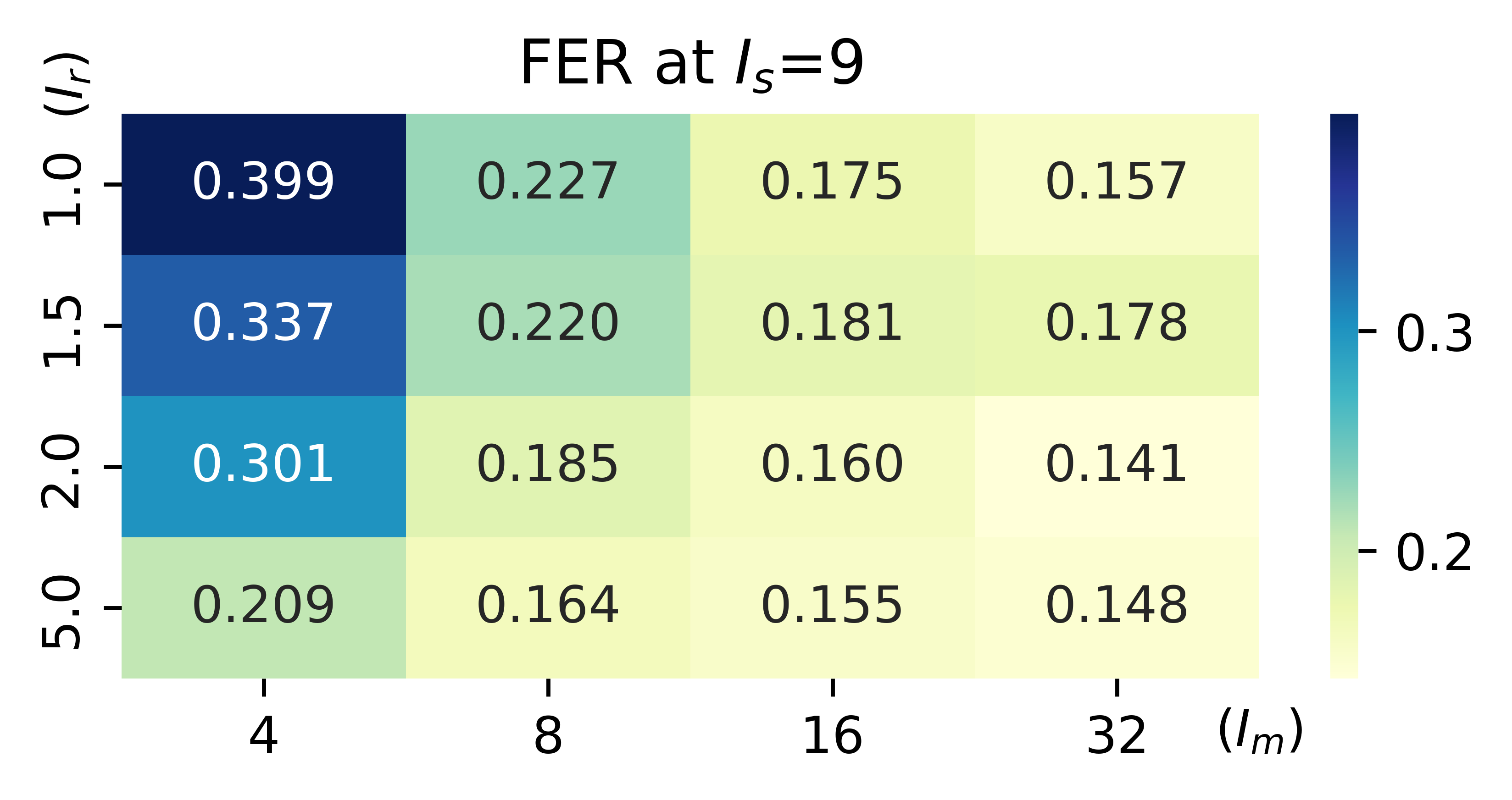}
		\subcaption{BCH (63,30) Code}
	\end{minipage}
	\hfill
	\begin{minipage}[t]{0.24\textwidth}
		\centering
		\includegraphics[width=\textwidth]{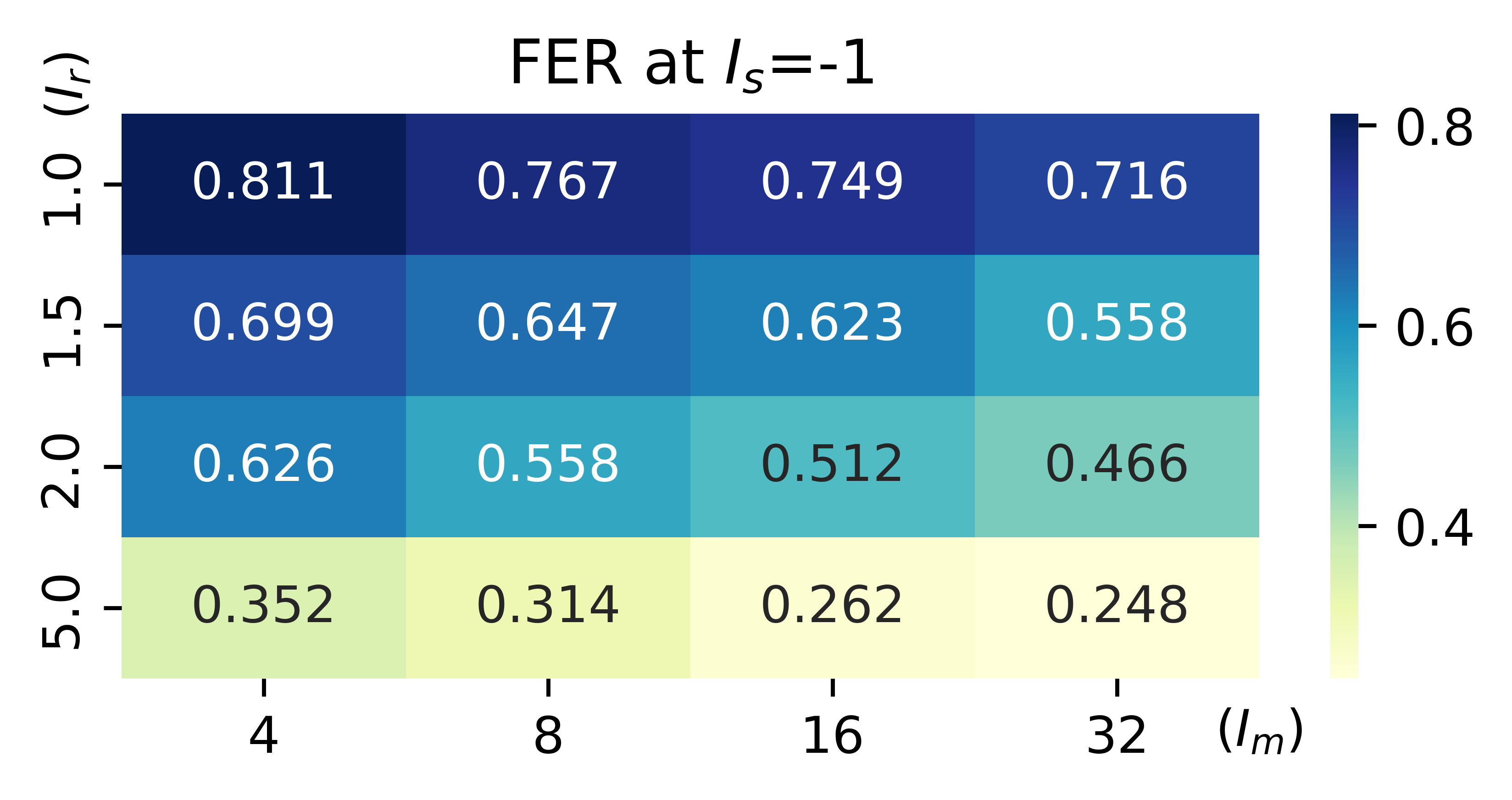}
		\vspace{0.2cm}
		\includegraphics[width=\textwidth]{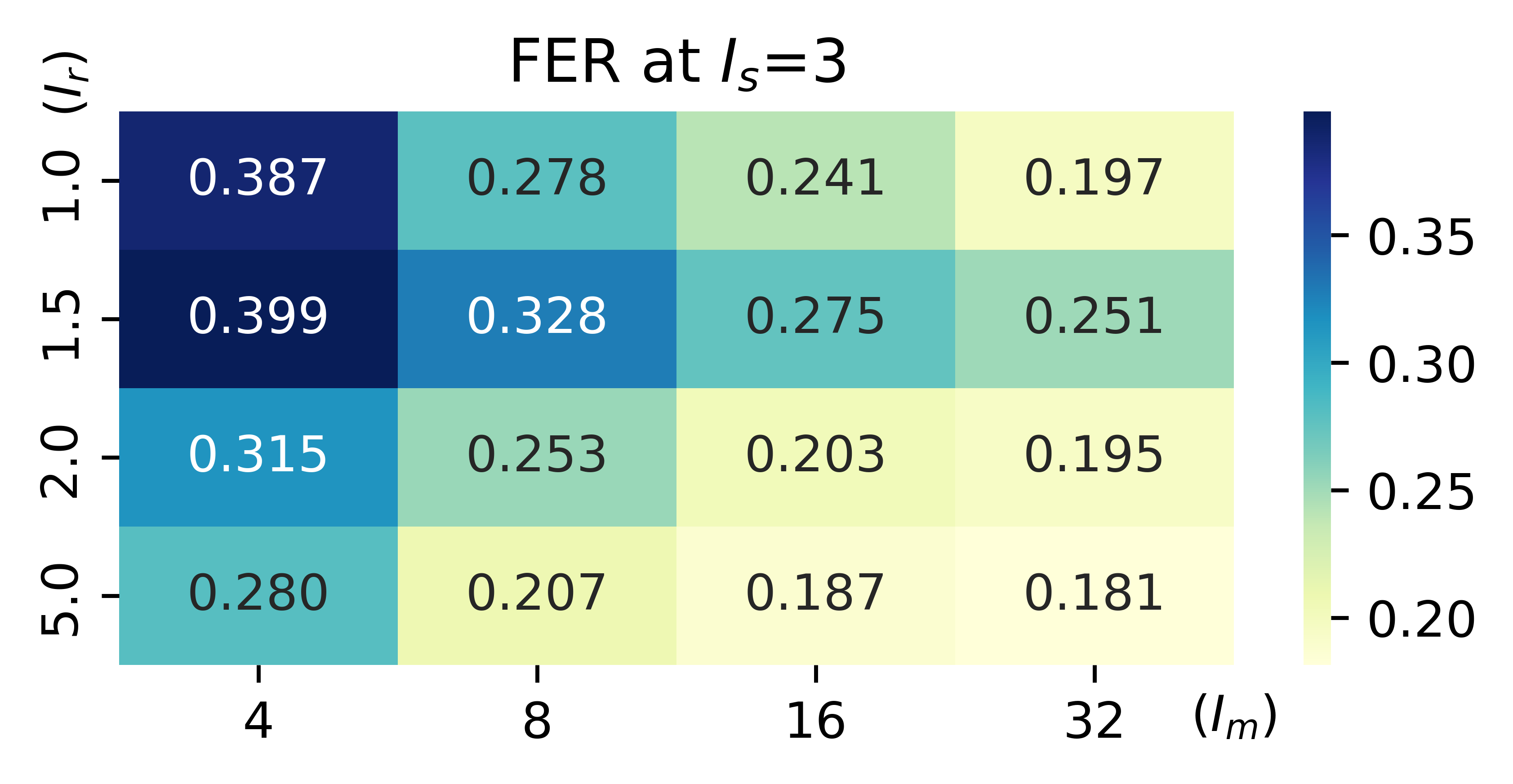}
		\vspace{0.2cm}
		\includegraphics[width=\textwidth]{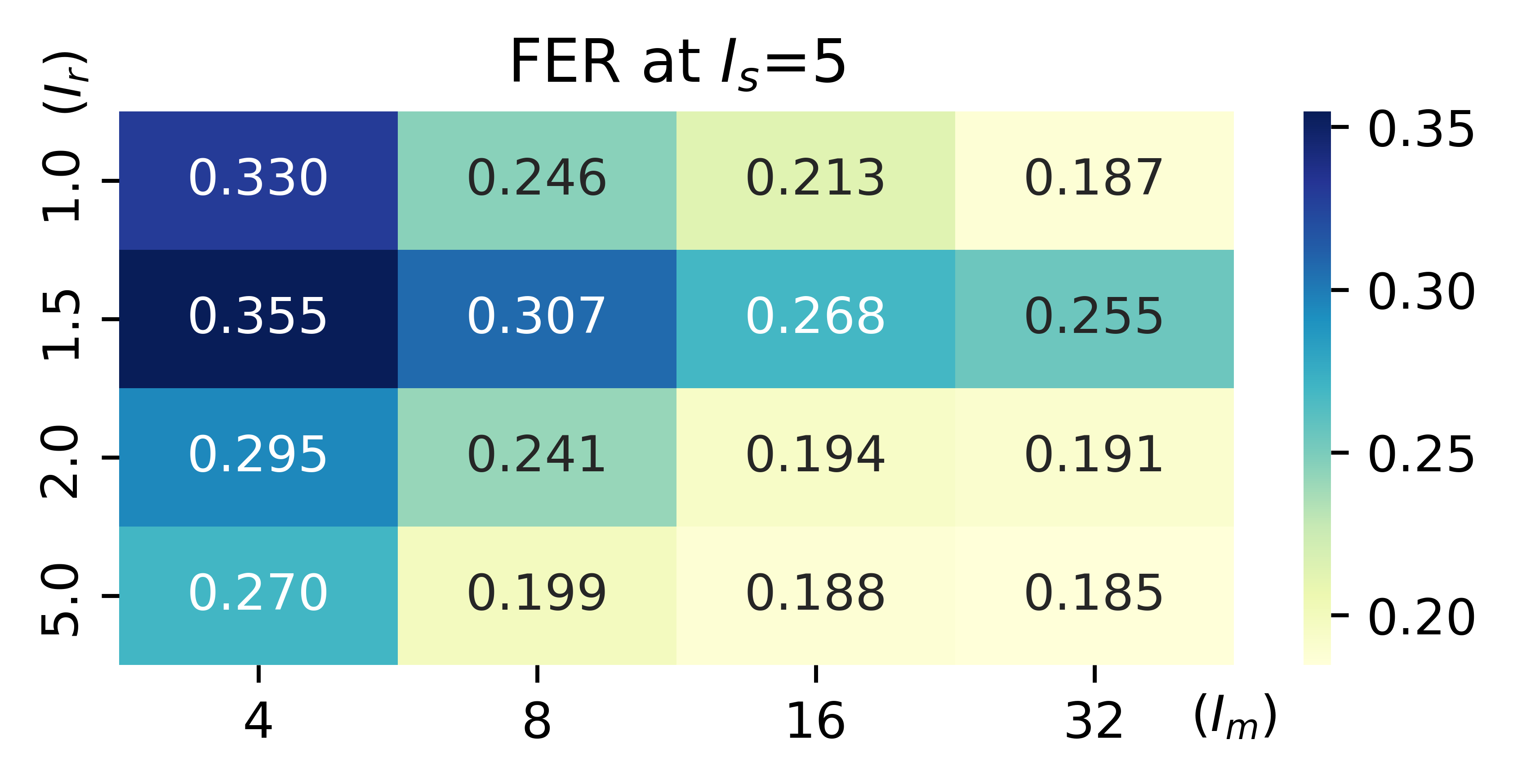}
		\subcaption{RS (15,7) Code}
	\end{minipage}  
	\caption{Heatmaps showing the strong dependence of FER on $(I_s,I_r,I_m)$ for BCH (63,30) and RS (15,7) codes at SNR = 2.0 dB.}
	\label{fig:comparison}
\end{figure}

To investigate how $(I_s,I_r,I_m)$ jointly affect FER performance, we tested the BCH (63,30) and RS (15,7) codes at SNR = 2.0 dB, assuming that $\alpha$ has been optimized for each chosen $(I_s,I_r,I_m)$. As shown in Fig.~\ref{fig:comparison}, several key trends emerge. First, FER generally improves with increasing $I_s$. A sharp performance gain occurs when $I_s$ increases from $-1$ to positive values, confirming that random cyclic shifts effectively mitigate short-cycle correlations in the Tanner graph. Second, not all high-valued combinations of $(I_s,I_r,I_m)$ necessarily yield the best FER performance, implicitly suggesting a trade-off among the competing effects of the respective hyperparameter settings. Third, for a fixed $I_s$, increasing $I_r$ gradually improves FER until saturation, reflecting the trade-off between the conflicting effects of added redundancy and newly introduced short cycles in the resulting $\mathbf{H}_o$. Finally, increasing $I_m$ generally enhances FER but with diminishing returns, highlighting the need to balance complexity and performance.

These trends generalize across block lengths and code rates. In particular, higher-rate BCH and RS codes converge more quickly with fewer iterations. Accordingly, in the following sections, for codes with a rate above one-half, we fix $I_m = 4$ for block lengths below 100 and $I_m = 8$ otherwise. For simplicity, we set $I_r = 2$ across all BCH and RS codes, with $I_s = 9$ for BCH codes and $I_s = 5$ for RS codes in the following discussions.

Assuming the worst-case NMS complexity scales as $|I_s| \cdot I_r \cdot I_m$, we define a complexity ratio $C_r$ relative to the baseline $(I_s, I_r, I_m) = (-1,1,4)$. For the BCH (63,30) code, compared to the optimal FER of 0.141 achieved by $(9,2,32)$ with $C_r = 144$, the configuration $(9,2,8)$ achieves an FER of 0.185 with $C_r = 36$, thus one quarter of the complexity of the former at the expense of a 0.044 FER loss. For the RS (15,7) binary image, $(3,2,16)$ achieves an FER of 0.203 with $C_r = 24$, versus 0.181 for $(3,5,32)$ with $C_r = 120$, a similarly small FER loss in exchange for significant complexity savings. These comparisons justify proper tuning of $(I_s, I_r, I_m)$ to seek a favorable trade-off between performance and complexity.
\subsection{Decoding Path Update}

In OSD, the decoding path is defined as the ordered set of TEPs of length $l_t$. Different decoding paths may yield marked variations in FER and complexity even for the same $l_t$.

In the original \uc{ALMLT}~\cite{kabat2007new}, the decoding path is derived from ordered statistics of all codeword bits under the assumptions of long block lengths and AWGN at a given SNR. In practice, however, the Gaussian elimination step in OSD introduces column swapping, and finite block length effects cause deviations between empirical means and \uc{ALMLT} estimates. Moreover, in the hybrid NMS-OSD architecture, OSD processes only NMS decoding failures, so the AWGN assumption no longer holds strictly.

\begin{figure}[htbp]
	\centering
	\begin{subfigure}[b]{0.24\textwidth}
		\centering
		\includegraphics[width=\textwidth]{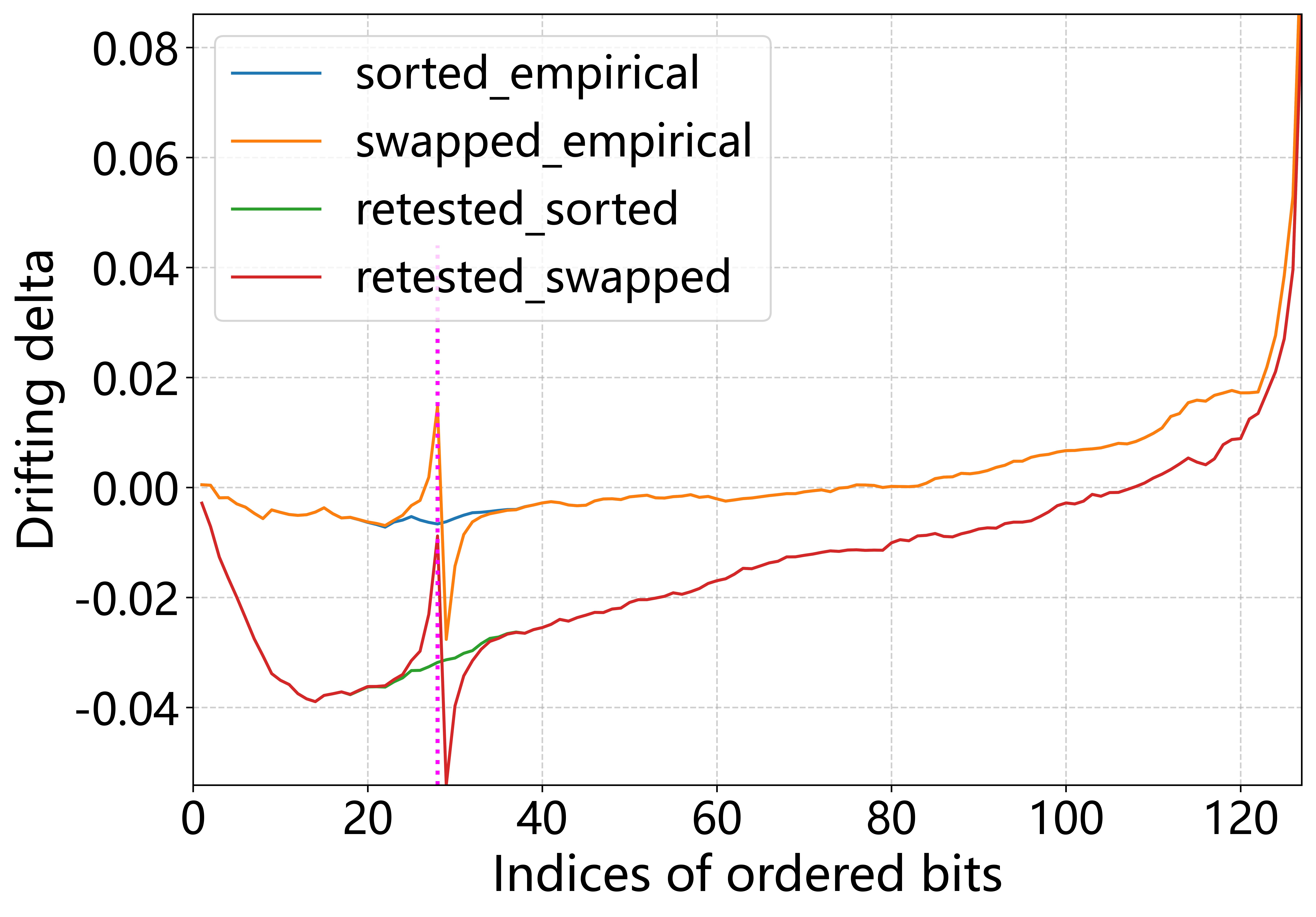}
		\caption{BCH (127,99) code}
		\label{fig:mean_reliability_3.5dB_BCH}
	\end{subfigure}
	\hfill
	\begin{subfigure}[b]{0.24\textwidth}
		\centering
		\includegraphics[width=\textwidth]{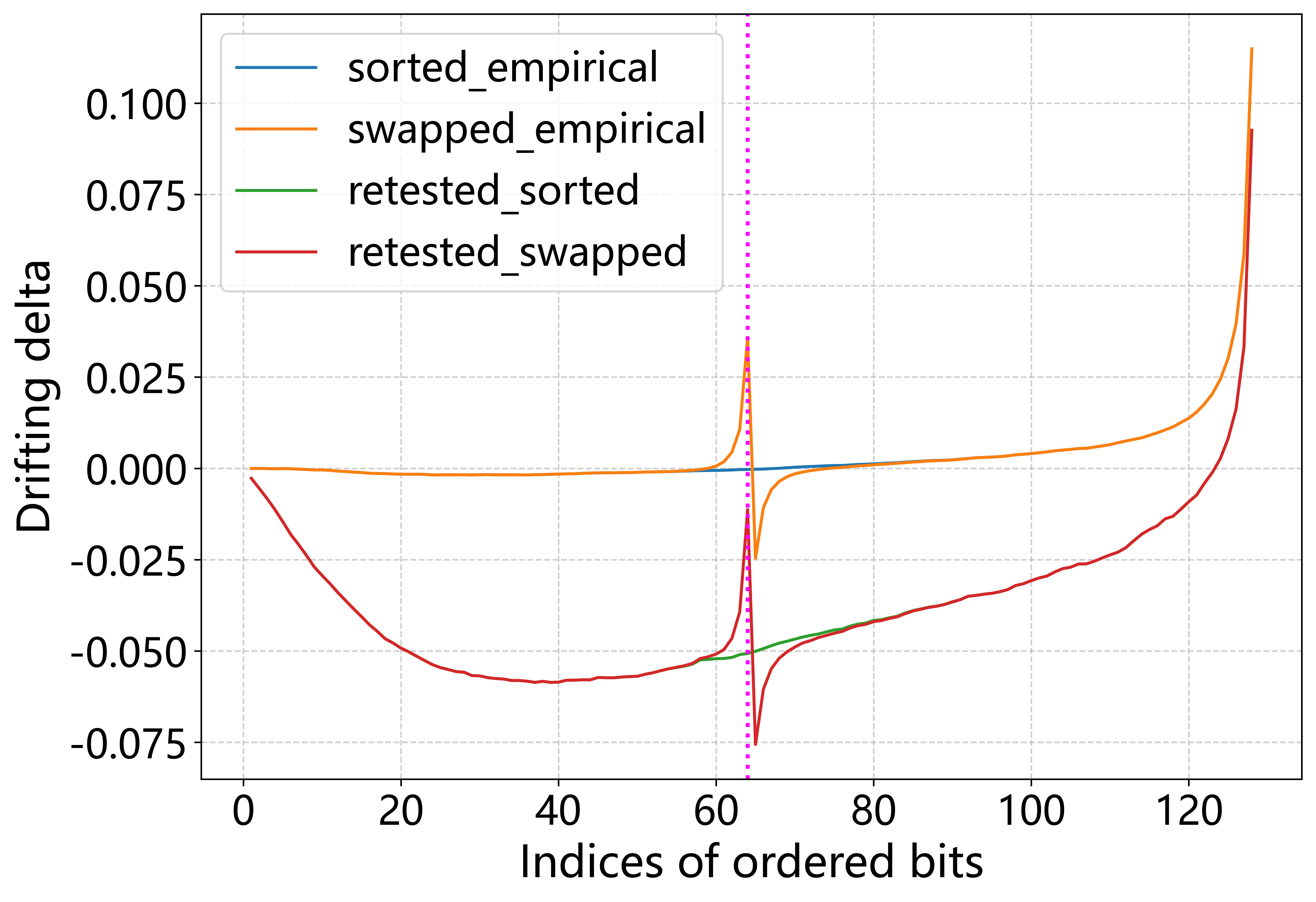}
		\caption{LDPC (128,64) code}
		\label{fig:mean_reliability_3.5dB_LDPC}
	\end{subfigure}
	\caption{Mean delta between empirical and \uc{ALMLT} calculations for BCH (127,99) and LDPC (128,64) codes at SNR = 3.5 dB. Vertical dotted lines mark MRB/LRB boundaries.}
	\label{fig:mean_reliability_3.5dB}
\end{figure}

We define the mean delta as the \uc{ALMLT} calculation subtracted from the empirical mean. The curves \textit{sorted\_empirical} and \textit{swapped\_empirical} represent the empirical statistics of the original received sequences, depending on whether the effect of Gaussian elimination-induced column swapping is taken into account. The curves \textit{retested\_sorted} and \textit{retested\_swapped} further restrict to NMS-failure cases. Ideally, mean deltas should approach zero, indicating close agreement with the \uc{ALMLT} calculation.

As shown in Fig.~\ref{fig:mean_reliability_3.5dB}, for both BCH (127,99) and LDPC (128,64) codes, \textit{sorted\_empirical} aligns closely with \uc{ALMLT} except for a few bits in the rear codeword section. In contrast, \textit{swapped\_empirical} exhibits sharp deviations near the MRB/LRB boundaries (bit indices 28 and 64, respectively). However, \textit{retested\_sorted} and \textit{retested\_swapped} deviate across all bit indices, with larger discrepancies in the LRB. Similar phenomena are observed for other SNRs as well as for RS codes, motivating the incorporation of empirical evidence to enhance the fixed \uc{ALMLT} ordering.

The enhanced \uc{ALMLT} proceeds as follows. First, the foremost $l_w = \beta \cdot l_t$ TEPs are extracted by the \uc{ALMLT} algorithm, where $\beta$ is a relaxing factor determined empirically. The decoding path comprises the first $l_t$ of the $l_w$ TEPs, while the remaining $l_w - l_t$ TEPs serve as a buffer that preserves the opportunity to swap with TEPs in the decoding path. Each member of the $l_w$ TEPs is assigned a counter initialized by its index. Next, upon receiving the first batch of samples, the counters are updated by subtracting the occurrences of true error patterns, and the $l_w$ TEPs are resorted according to the updated counts. As additional batches arrive, the counter evaluations are updated consecutively, allowing the decoding path to adapt online. In practice, estimated OSD error patterns function as the true ones to update the related counters.

In this enhanced scheme, \uc{ALMLT} provides a fairly strong starting point, while empirical updates naturally adjust the rankings of TEPs to achieve instantaneous adaptiveness to channel variation. In the long run, this approach is expected to reduce the average number of TEPs required in OSD due to the forward movement of frequently emerging true error patterns underestimated by \uc{ALMLT}, and to improve FER  because true error patterns located in the buffer have the chance to enter the decoding path through this update mechanism.

\setlength{\tabcolsep}{2pt}    
\begin{table}[htp]
	\caption{Average number of TEPs before meeting the true error pattern under various decoding path schemes of OSD. Parameters: $(l_t,\beta)=(10^3,10)$ for BCH/RS codes and $(10^4,4)$ for the LDPC code.}
	\label{tab:averag_tep_ror}
	\resizebox{\columnwidth}{!}{%
		\begin{tabular}{|cc|c|c|c|}
			\hline
			\multicolumn{2}{|c|}{Codes} &
			\uc{\textit{CVT}} &
			\uc{\textit{ALMLT}} &
			\uc{\textit{ALMLT\_MS}} \\ \hline
			\multicolumn{1}{|c|}{} &
			\begin{tabular}[c]{@{}c@{}}Plain\end{tabular} &
			{\color{cyan!50!blue} \begin{tabular}[c]{@{}c@{}}77.76/40.71/\\ 21.28/10.68\end{tabular}} &
			{\color{cyan!50!blue} \begin{tabular}[c]{@{}c@{}}54.61/27.03/\\ 13.92/7.38\end{tabular}} &
			{\color{cyan!50!blue} \begin{tabular}[c]{@{}c@{}}54.58/27.02/\\ 13.91/7.38\end{tabular}} \\ \cline{2-5} 
			\multicolumn{1}{|c|}{\multirow{-3}{*}{\begin{tabular}[c]{@{}c@{}}BCH (127,99)\\ SNR=3.0/3.5/\\ 4.0/4.5 dB\end{tabular}}} &
			\begin{tabular}[c]{@{}c@{}} DIA\end{tabular} &
			{\color{blue!80!black} \begin{tabular}[c]{@{}c@{}}76.64/39.41/\\ 19.76/9.48\end{tabular}} &
			{\color{blue!80!black} \begin{tabular}[c]{@{}c@{}}53.72/26.22/\\ 12.87/6.41\end{tabular}} &
			{\color{blue!80!black} \begin{tabular}[c]{@{}c@{}}53.70/26.21/\\ 12.88/6.41\end{tabular}} \\ \hline
			\multicolumn{1}{|c|}{} &
			\begin{tabular}[c]{@{}c@{}}Plain\end{tabular} &
			{\color{cyan!50!blue} \begin{tabular}[c]{@{}c@{}}124.63/66.27/\\ 32.54/14.42\end{tabular}} &
			{\color{cyan!50!blue} \begin{tabular}[c]{@{}c@{}}90.43/44.95/\\ 21.38/9.51\end{tabular}} &
			{\color{cyan!50!blue} \begin{tabular}[c]{@{}c@{}}90.40/44.93/\\ 21.39/9.50\end{tabular}} \\ \cline{2-5} 
			\multicolumn{1}{|c|}{\multirow{-3}{*}{\begin{tabular}[c]{@{}c@{}}RS (31,25)\\ SNR=3.0/3.5/\\ 4.0/4.5 dB\end{tabular}}} &
			\begin{tabular}[c]{@{}c@{}} DIA\end{tabular} &
			{\color{blue!80!black} \begin{tabular}[c]{@{}c@{}}123.42/64.6/\\ 30.56/13.35\end{tabular}} &
			{\color{blue!80!black} \begin{tabular}[c]{@{}c@{}}89.08/43.99/\\ 19.75/8.82\end{tabular}} &
			{\color{blue!80!black} \begin{tabular}[c]{@{}c@{}}89.04/43.97/\\ 19.76/8.81\end{tabular}} \\ \hline
			\multicolumn{1}{|c|}{} &
			\begin{tabular}[c]{@{}c@{}}Plain\end{tabular} &
			{\color{cyan!50!blue} \begin{tabular}[c]{@{}c@{}}799.68/485.72/\\ 325.16/191.96\end{tabular}} &
			{\color{cyan!50!blue} \begin{tabular}[c]{@{}c@{}}604.87/357.77/\\ 231.66/135.65\end{tabular}} &
			{\color{cyan!50!blue} \begin{tabular}[c]{@{}c@{}}604.89/357.79/\\ 231.71/135.70\end{tabular}} \\ \cline{2-5} 
			\multicolumn{1}{|c|}{\multirow{-3}{*}{\begin{tabular}[c]{@{}c@{}}LDPC (128,64)\\ SNR=2.0/2.5/\\ 3.0/3.5 dB\end{tabular}}} &
			\begin{tabular}[c]{@{}c@{}} DIA\end{tabular} &
			{\color{blue!80!black} \begin{tabular}[c]{@{}c@{}}426.2/173.89/\\ 86.48/39.92\end{tabular}} &
			{\color{blue!80!black} \begin{tabular}[c]{@{}c@{}}321.46/125.78/\\ 60.28/28.09\end{tabular}} &
			{\color{blue!80!black} \begin{tabular}[c]{@{}c@{}}321.46/125.79/\\ 60.28/28.09\end{tabular}} \\ \hline
		\end{tabular}
	}
\end{table}

Denote \uc{\textit{ALMLT\_MS}} as a snapshot of the enhanced \uc{ALMLT} refreshed by the decoding failures of NMS decoding from the initial $10^5$ received channel sequences, and \uc{\textit{CVT}} as the intercepted $l_t$ TEPs at the head of the fixed conventional TEP list~\cite{Fossorier1995}. The \uc{ALMLT} decoding path for BCH and RS codes is calculated at SNR = 3.5 dB with $(l_t,\beta) = (10^3, 10)$; for the LDPC code, it is calculated at SNR = 3.0 dB with $(10^4, 4)$. Table~\ref{tab:averag_tep_ror} reports the average number of TEPs before meeting the true error pattern for the concerned decoding paths.

Focusing on the rows labeled 'Plain' (i.e., without DIA), a consistent reduction in the average number of TEPs is observed for \uc{\textit{ALMLT}} over \uc{\textit{CVT}} across all codes and SNR points. This indicates that mean-based TEP ordering substantially surpasses Hamming weight-based TEP ordering. However, the marginal gain for BCH and RS codes and the slight degradation for the LDPC code when comparing \uc{\textit{ALMLT\_MS}} with \uc{\textit{ALMLT}} suggest the following. First, \uc{ALMLT} already presents a sufficiently optimized TEP ordering, leaving little room for further improvement — especially since the ordering of the foremost TEP list perfectly aligns with empirical data. Second, due to the relatively small sample sizes involved in updating, a single snapshot struggles to reflect the long-term effect of large-$l_t$ settings. As more samples accumulate and the counters for each TEP reach statistical stability, \uc{\textit{ALMLT\_MS}} is expected to surpass both baselines for all codes, since the aforementioned mean delta can be largely compensated by incorporating historical data to refresh the TEP ordering.
\subsection{Refined Reliability Estimates via DIA Model}
In prior work~\cite{li2024boosting}, the DIA model was shown to improve reliability estimates for LDPC codeword bits. Herein, we elaborate on it for LDPC codes and extend its application to BCH and RS codes.

\begin{figure}[htbp]
	\centering
	\begin{subfigure}[b]{0.24\textwidth}
		\centering
		\includegraphics[width=\textwidth]{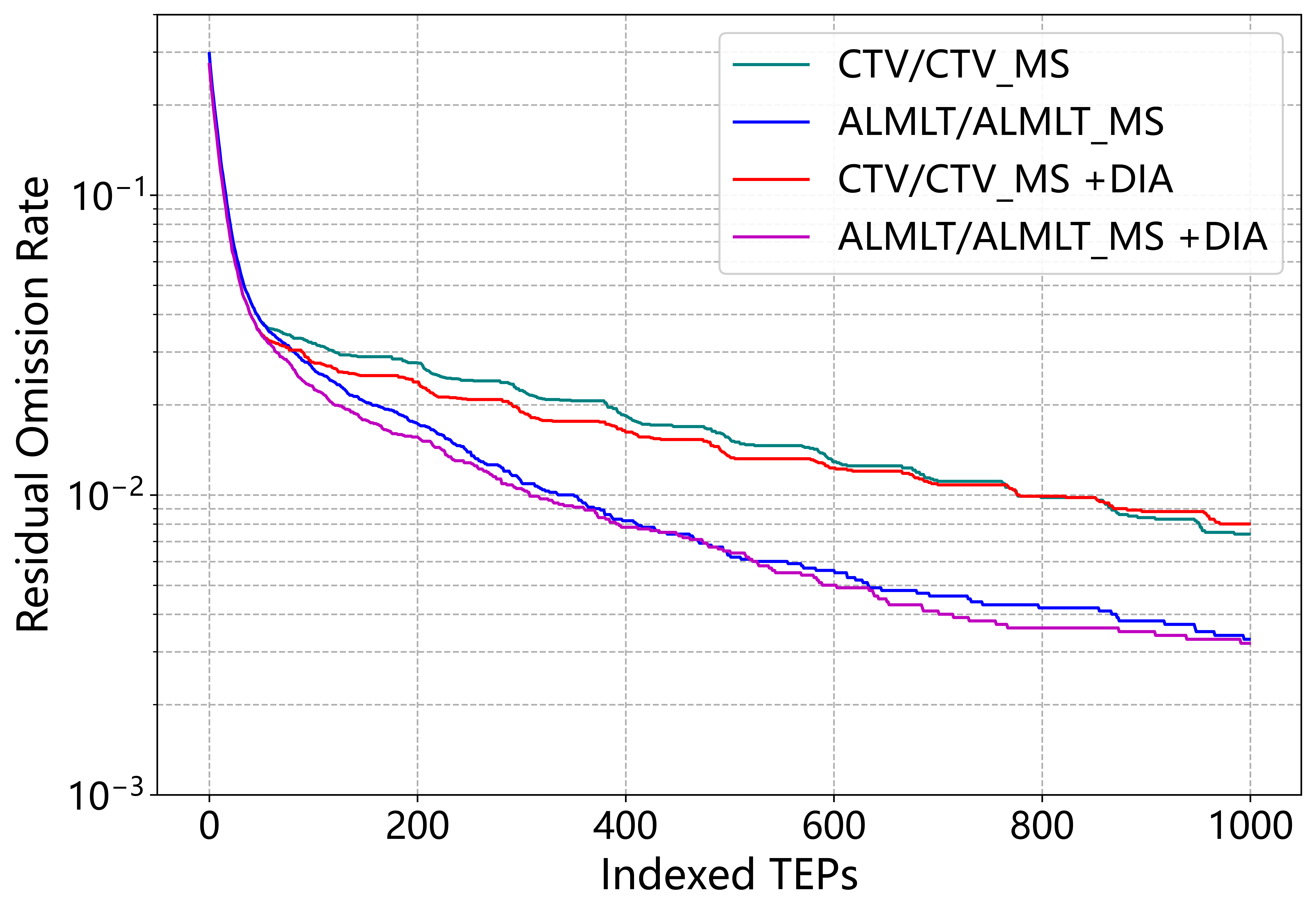}
		\caption{ROR for BCH code}
		\label{fig:bch_4.0dB_ror}
	\end{subfigure}
	\hfill
	\begin{subfigure}[b]{0.24\textwidth}
		\centering
		\includegraphics[width=\textwidth]{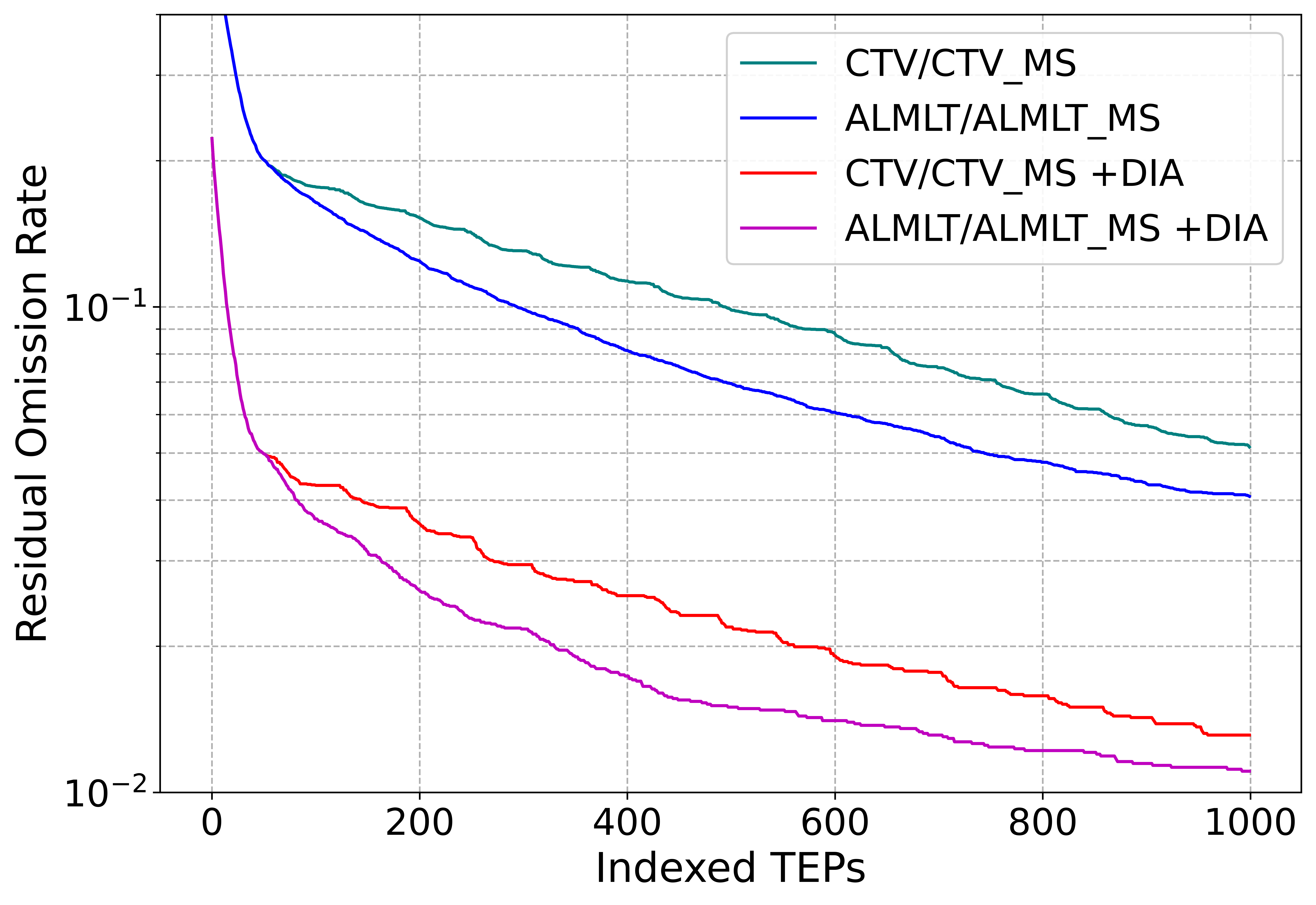}
		\caption{ROR for LDPC code}
		\label{fig:ldpc_3.0dB_ror}
	\end{subfigure}
	\caption{Residual omission rate for decoding failures of NMS with $(I_s,I_r,I_m)=(9,2,8)$ for BCH (127,99) and with $I_m=10$ for LDPC (128,64) codes at SNR = 4.0 and 3.0 dB, respectively, under various decoding paths.}
	\label{fig:ror_bch_ldpc}
\end{figure}

We define the residual omission rate (ROR), an FER-like metric, as the fraction of NMS decoding failures that remain unresolved after the $i$-th TEP scanning in the decoding path of OSD. ROR results are shown in Fig.~\ref{fig:ror_bch_ldpc} for four basic decoding paths: \uc{\textit{CVT}}, its empirically-enhanced variant \uc{\textit{CVT\_MS}}, \uc{\textit{ALMLT}}, and \uc{\textit{ALMLT\_MS}}. Due to limited empirical data, the \uc{\textit{CVT}} and \uc{\textit{CVT\_MS}} curves are nearly indistinguishable, as are \uc{\textit{ALMLT}} and \uc{\textit{ALMLT\_MS}}; they are therefore roughly grouped as \uc{\textit{CVT/CVT\_MS}} and \uc{\textit{ALMLT/ALMLT\_MS}}, respectively. The addition of DIA yields two further distinct ROR curves.

For BCH codes, Fig.~\ref{fig:bch_4.0dB_ror} shows that the DIA provides marginal but consistent gains. A closer inspection of this minor effect is presented in Table~\ref{tab:tep_distribution_bch}. In the first column, DIA trained at SNR = 4.0 dB consistently shifts high Hamming-weight true error patterns toward lower weights across all SNR points. In the second column, true error patterns are moved forward into the head blocks of the decoding path. Thus, the empirical data indicate that both FER and $I_{at}$ measurements benefit from DIA, whether given a fixed order-$p$ TEP list or a TEP list of fixed $l_t$ when constructing the decoding path.

\begin{table}[htbp]
	\caption{Impact of DIA trained at SNR = 4.0 dB for the decoding failures of NMS with the setting $(I_s,I_r,I_m)=(9,2,8)$ for BCH (127,99) code, in terms of number of erroneous bits in MRB (first column) and hits in the indexed blocks in decoding path (second column) for $l_t=10^3$, where '-1' denotes hits overflowed beyond $l_t$.}
	\label{tab:tep_distribution_bch}
	\resizebox{\columnwidth}{!}{%
		\begin{tabular}{|cc|c|c|}
			\hline
			\multicolumn{2}{|c|}{SNR}               & Distribution of erroneous bits in MRB                         & Hits by indexed blocks ($b_s=100$)                           \\ \hline
			\multicolumn{1}{|c|}{\multirow{2}{*}{2.5dB}} & Plain                      & \{0: 2129, 1: 1397, 2: 421, 3: 65,...\} & \{0: 3591, 1: 98, -1: 96, 2: 68,...\}      \\ \cline{2-4} 
			\multicolumn{1}{|c|}{}                       & DIA                        & \{\color{blue!80!black}\textbf{0: 2258}, 1: 1272, 2: 415, 3: 66,...\} & \{\color{blue!80!black}\textbf{0: 3596}, -1: 99, 1: 97, 2: 74,...\}      \\ \hline
			\multicolumn{1}{|c|}{\multirow{2}{*}{3.0dB}} & \multicolumn{1}{l|}{Plain} & \{0: 6199, 1: 3063, 2: 654, 3: 84,...\} & \{0: 9369, 1: 179, -1: 129,...\}           \\ \cline{2-4} 
			\multicolumn{1}{|c|}{}                       & DIA                        & \{\color{blue!80!black}\textbf{0: 6614}, 1: 2748, 2: 566, 3: 73,...\} & \{\color{blue!80!black}\textbf{0: 9456}, 1: 155, -1: 110,...\}           \\ \hline
			\multicolumn{1}{|c|}{\multirow{2}{*}{3.5dB}} & \multicolumn{1}{l|}{Plain} & \{0: 7063, 1: 2519, 2: 382, 3: 37,...\} & \{0: 9658, 1: 108, 2: 59, -1: 53,...\} \\ \cline{2-4} 
			\multicolumn{1}{|c|}{}                       & DIA                        & \{\color{blue!80!black}\textbf{0: 7588}, 1: 2099, 2: 279, 3: 35,...\} & \{\color{blue!80!black}\textbf{0: 9732}, 1: 78, -1: 47,...\}             \\ \hline
			\multicolumn{1}{|c|}{\multirow{2}{*}{4.0dB}} & \multicolumn{1}{l|}{Plain} & \{0: 12918, 1: 3313, 2: 314,...\}       & \{0: 16291, 1: 83, 2: 51, -1: 40,...\}     \\ \cline{2-4} 
			\multicolumn{1}{|c|}{}                       & DIA                        & \{\color{blue!80!black}\textbf{0: 13800}, 1: 2526, 2: 235,...\}       & \{\color{blue!80!black}\textbf{0: 16365}, 1: 78, 2: 37,..., -1: 16,...\} \\ \hline
			\multicolumn{1}{|c|}{\multirow{2}{*}{4.5dB}} & \multicolumn{1}{l|}{Plain} & \{0: 5939, 1: 1038, 2: 87,...\}         & \{0: 6988, 1: 36, 2: 15,..., -1: 7,...\}   \\ \cline{2-4} 
			\multicolumn{1}{|c|}{}                       & DIA                        & \{\color{blue!80!black}\textbf{0: 6282}, 1: 740, 2: 45,...\}          & \{\color{blue!80!black}\textbf{0: 7030}, 1: 17, ..., -1: 3,...\}         \\ \hline
		\end{tabular}
	}
\end{table}
For LDPC codes, the improvement is much more pronounced: Fig.~\ref{fig:ldpc_3.0dB_ror} shows that for both \uc{\textit{CTV}} and \uc{\textit{ALMLT}}, the DIA is expected to significantly reduce the FER level observed in its absence.

Indeed, the rows labelled as 'DIA' in Table~\ref{tab:averag_tep_ror} of the previous section corroborate these improvements, showing that the application of DIA reduces the average TEP count (and thus $I_{at}$) across all codes. With sufficient samples, the \uc{\textit{CVT\_MS}} and \uc{\textit{ALMLT\_MS}} curves are expected to converge, reflecting the overriding influence of statistical evidence over the initial TEP ordering.

An optimal decoding path ideally produces a smoothly convex ROR curve, corresponding to monotonically decreasing contributions from successive TEPs. Fig.~\ref{fig:ror_bch_ldpc} indicates that \uc{\textit{CVT\_MS}} requires extensive sample collection and many TEP reorderings to smooth fluctuations, whereas \uc{\textit{ALMLT\_MS}} achieves comparable smoothness with far fewer samples, highlighting \uc{ALMLT} as an effective initialization strategy.
\begin{table}[htbp]
	\caption{Impact of the DIA model on OSD decoding with an ALMLT decoding path of length $l_t=10^3$ for LDPC (128,64) and BCH (127,64) codes, trained at SNR = 3.0 dB and SNR = 3.5 dB, respectively.}
	\label{tab:dia_effect_fer}
	\resizebox{\columnwidth}{!}{
	\begin{tabular}{|c|c|cl|}
		\hline
		&       & \multicolumn{2}{c|}{}                                                                                     \\
		\multirow{-2}{*}{Codes} &
		\multirow{-2}{*}{Abs / Avail} &
		\multicolumn{2}{c|}{\multirow{-2}{*}{(SNR,FER) OSD post-processing of NMS decoding failures}} \\ \hline
		& Plain & \multicolumn{2}{c|}{{[}(1.5, 0.16917), (2.0, 0.10526), (2.5, 0.06833), (3.0, 0.04578){]}} \\ \cline{2-4} 
		\multirow{-2}{*}{LDPC(128,64)} &
		DIA &
		\multicolumn{2}{c|}{{\color{blue!80!black}\textbf {[}(1.5, 0.11824), (2.0, 0.058), (2.5, 0.02968), (3.0, 0.01639){]}}} \\ \hline
		& Plain & \multicolumn{2}{c|}{{[} (2.5, 0.03908), (3.0, 0.01955), (3.5, 0.0076), (4.0, 0.00277){]}}    \\ \cline{2-4} 
		\multirow{-2}{*}{BCH(127,64)} &
		DIA &
		\multicolumn{2}{c|}{{\color{blue!80!black}\textbf {[} (2.5, 0.03569), (3.0, 0.01715), (3.5, 0.0056), (4.0, 0.00188){]}}} \\ \hline
	\end{tabular}
	}
\end{table}

To isolate the impact of the DIA model on the FER of OSD, and assuming the ALMLT ordering of TEPs along a decoding path of length $l_t=10^3$, empirical data for LDPC (128,64) and BCH (127,64) codes are presented in Table~\ref{tab:dia_effect_fer}. It is observed that the DIA model generalizes well across all SNR regions of interest, despite being trained at a specific SNR (3.0 dB for LDPC and 3.5 dB for BCH). It is also found that the improvement in OSD performance on NMS decoding failures is considerably more pronounced for LDPC codes than for BCH codes, which corroborates the findings in Table~\ref{tab:tep_distribution_bch}.

More advanced NN structures beyond CNNs were attempted but yielded little additional gain. This suggests that the message updating mechanism of ENMS, designed for BCH and RS codes, does not produce the same informative posterior LLR trajectories derived from standard NMS that benefit the DIA model for LDPC codes.
\subsection{Revised SWA-Based OSD}

When the length $l_t$ of a decoding path must be set sufficiently large to approach near-ML performance for longer block codes, it is crucial to terminate TEP traversal once the true error pattern emerges to reduce computational load and decoding latency. Given the decoding path \textit{ALMLT\_MS} and a trained SWA model, the revised SWA-based OSD proceeds as follows.

First, prepare the inputs for the SWA model. Fetch the first $w_t$ blocks of TEPs from \textit{ALMLT\_MS}, each of uniform size $b_s$. Then extract and sort the $w_t$ minima with respect to the distance to the received sequence. These minima, together with the current window index (initialized to zero), form the input window of the SWA model. The SWA model then determines whether early stopping should occur, with the global minimum $g_m$ initialized as the smallest among the $w_t$ minima.

Next, the window slides along the decoding path by one block at a time: the newest block is admitted, and the oldest block is discarded. The SWA model is triggered when the minimum of the newest block is less than $g_m$, which then updates $g_m$. This process continues until the early stopping criterion is met or the decoding path is fully traversed, with the final estimate corresponding to the TEP associated with distance $g_m$.

Concerning the early stopping criterion, we define a soft margin $s_m$ as the difference between the binary probabilities of the SWA model outputs. A larger $s_m$ enforces stricter early-stopping conditions, reducing FER degradation but increasing the average TEP count $I_{at}$. Thus, tuning $s_m$ can directly balance performance and complexity.

\setlength{\tabcolsep}{6pt}
\begin{table}[htp]
\caption{FER and $I_{at}$ for SWA-based OSD with DIA, applied to NMS decoding failures of BCH (127,99) with $l_t=10^3$, RS (31,25) with $l_t=3\times10^3$, and LDPC (128,64) with $l_t=10^4$.}
	\label{tab:fer_average_tep_codes}
	\resizebox{\columnwidth}{!}{%
		\begin{tabular}{|c|c|ccc|}
			\hline
			& & \multicolumn{3}{c|}{FER/$I_{at}$} \\ \cline{3-5} 
			\multirow{-2}{*}{Codes} & \multirow{-2}{*}{$b_s$} & \multicolumn{1}{c|}{$s_m=0.2$} & \multicolumn{1}{c|}{$s_m=0.9$} & $s_m=1.0$ \\ \hline
			& 10 & \multicolumn{1}{c|}{{\color{blue!80!black} 2.9e-2/82.1}} & \multicolumn{1}{c|}{{\color{blue!80!black} 2.3e-2/167.5}} & {\color{blue!80!black} 2.2e-2/1e3} \\ \cline{2-2} 
			\multirow{-2}{*}{\begin{tabular}[c]{@{}c@{}}BCH (127,99)\\ SNR=3.5dB\end{tabular}} & 100 & \multicolumn{1}{c|}{{\color{blue!80!black} 2.3e-2/515.3}} & \multicolumn{1}{c|}{{\color{blue!80!black} 2.2e-2/558.4}} & {\color{blue!80!black} 2.2e-2/1e3} \\ \hline
			& 10 & \multicolumn{1}{c|}{{\color{blue!80!black} 3.3e-2/272.5}} & \multicolumn{1}{c|}{{\color{blue!80!black} 3.0e-2/809.2}} & {\color{blue!80!black} 3.0e-2/3e3} \\ \cline{2-2} 
			\multirow{-2}{*}{\begin{tabular}[c]{@{}c@{}}RS (31,25)\\ SNR=3.5dB\end{tabular}} & 100 & \multicolumn{1}{c|}{{\color{blue!80!black} 3.0e-2/682.9}} & \multicolumn{1}{c|}{{\color{blue!80!black} 3.0e-2/1030.2}} & {\color{blue!80!black} 3.0e-2/3e3} \\ \hline
			& 10 & \multicolumn{1}{c|}{{\color{blue!80!black} 1.5e-2/419.3}} & \multicolumn{1}{c|}{{\color{blue!80!black} 9.9e-3/653.8}} & {\color{blue!80!black} 8.4e-3/1e4} \\ \cline{2-2} 
			\multirow{-2}{*}{\begin{tabular}[c]{@{}c@{}}LDPC (128,64)\\ SNR=2.5dB\end{tabular}} & 100 & \multicolumn{1}{c|}{{\color{blue!80!black} 8.8e-3/808.2}} & \multicolumn{1}{c|}{{\color{blue!80!black} 8.4e-3/1263.1}} & {\color{blue!80!black} 8.4e-3/1e4} \\ \hline
		\end{tabular}
	}
\end{table}

Table~\ref{tab:fer_average_tep_codes} reports FER and $I_{at}$ for BCH, RS, and LDPC codes under appropriately chosen parameters $(l_t,\beta,w_t,b_s)=(10^3(10^4),10(4),5,b_s)$ for each case. Increasing $b_s$ allows more TEPs to be processed in parallel, thereby reducing time complexity but raising $I_{at}$. For instance, for the BCH (127,99) code at $s_m=0.9$, setting $b_s=10$ lowers $I_{at}$ to about one-third of the value obtained with $b_s=100$, at the expense of additional window slidings. Importantly, the average number of SWA calls consistently remains between 1 and 1.5, since most TEP traversals shift the window forward without activating SWA.

The data in Table~\ref{tab:fer_average_tep_codes} further reveal a tradeoff: a larger $s_m$ increases $I_{at}$ but reduces FER. The baseline case $s_m=1.0$ disables SWA, forcing all TEPs to be processed in parallel. This achieves the best FER but also incurs the highest $I_{at}$, making it practical only for short or moderate-length decoding paths. Conversely, a smaller $b_s$ minimizes $I_{at}$ but causes frequent window slidings and amplifies the sensitivity of both FER and $I_{at}$ to $s_m$.

A joint examination of Tables~\ref{tab:fer_average_tep_codes} and \ref{tab:averag_tep_ror} shows that, when $b_s=1$, the $I_{at}$ values in Table~\ref{tab:fer_average_tep_codes} are lower bounded by those in Table~\ref{tab:averag_tep_ror}. The latter table represents the ideal case where true error patterns are identified immediately upon appearance, whereas the former reflects the SWA model's ability to infer whether the global minimum has appeared in an incomplete candidate list. Due to the heavy penalty imposed on premature termination before encountering the true error pattern in the loss definition during training, the scenario of hastily stopping TEP scanning before the true error pattern emerges is nearly eliminated, while postponing until the end is also rare. Hence, a SWA model with an appropriately tuned $s_m$ can effectively regulate $I_{at}$ with negligible FER degradation, achieving a practical balance between performance and complexity.
\section{Experimental Results and Complexity Analysis}
\label{simulations}

\begin{figure*}[bt]
	\centering
	\begin{subfigure}[t]{0.32\textwidth}
		\begin{tikzpicture}
    \begin{semilogyaxis}[
    	width=10cm,
    	height=7cm,
        scale = 0.6,
        xlabel={$E_b/N_0$(dB)},
        ylabel={FER},
        xmin=1.0, xmax=4.0,
        ymin=1e-3, ymax=1.0,
        xtick={1.0,1.5,2,2.5,...,4.0},
        legend pos = south west,
        ymajorgrids=true,
        xmajorgrids=true,
        grid style=dashed,
        legend style={font=\fontsize{6}{7}\selectfont,fill opacity=0.8,text opacity=1,legend columns=2},
        ]
%0 plot for NMS (FER)
\addplot[
color=magenta,
mark=x,
very thin
]
coordinates {
(1.000, 0.70111)
(1.500, 0.59655)
(2.000, 0.46741)
(2.500, 0.33146)
(3.000, 0.22110)
(3.500, 0.13089)
(4.000, 0.07003)
};	
\addlegendentry{NMS(6)}

%1 plot for {NMS(6)+UDE(-1.0)+OSD($p=2$)}
\addplot[
color=blue,
mark= square,
very thin
]
coordinates {
(1.000, 0.517)
(1.500, 0.407)
(2.000, 0.280)
(2.500, 0.181)
(3.000, 0.103)
(3.500, 0.051)
(4.000, 0.023)
};	
\addlegendentry{N(6)-U(-1.0)-O(2)}
%2 plot for {NMS(6)+UDE(0.9)+OSD($p=2$)}
\addplot[
color=red,
mark=*,
very thin
]
coordinates {
(1.000, 0.512)
(1.500, 0.411)
(2.000, 0.276)
(2.500, 0.181)
(3.000, 0.106)
(3.500, 0.050)
(4.000, 0.022)
};	
\addlegendentry{N(6)-U(0.9)-O(2)}
%0.5plot for ML (FER)
\addplot[
color=black,
mark=diamond,
very thin
]
coordinates {
(1.00, 5.155e-01)
(1.50, 4.049e-01)
(2.00, 2.994e-01)
(2.50, 2.012e-01)
(3.00, 1.042e-01)
(3.50, 4.440e-02)
(4.00, 2.007e-02)
(4.50, 8.987e-03)
(5.00, 2.811e-03)
};	
\addlegendentry{ML\cite{helmling19}}
%5 plot for {FER_FP-UDE(-1.)}
\addplot[
color=teal,
mark=triangle,
dashed,
very thin
]
coordinates {
(1.000, 0.12319)
(1.500, 0.09702)
(2.000, 0.07472)
(2.500, 0.04954)
(3.000, 0.03115)
(3.500, 0.01637)
(4.000, 0.00741)
};
\addlegendentry{FER$_{\text{UDE(-1.0)}}$}
%0.5plot for {FER_FP-UDE(0.9)}
\addplot[
color=purple,
mark=triangle*,
dashed,
very thin
]
coordinates {
(1.000, 0.04295)
(1.500, 0.03882)
(2.000, 0.03024)
(2.500, 0.02212)
(3.000, 0.01643)
(3.500, 0.00837)
(4.000, 0.00419)
};	
\addlegendentry{FER$_{\text{UDE(0.9)}}$}
\end{semilogyaxis}
\end{tikzpicture}
\caption{LDPC (49,36) code}
\label{fer49_36_ldpc_ude}
	\end{subfigure}
	\hfill
	\begin{subfigure}[t]{0.32\textwidth}
		\phantom{000}
			\begin{tikzpicture}
		\begin{semilogyaxis}[
			width=10cm,
			height=7cm,
		scale = 0.6,	xlabel={$E_b/N_0$(dB)},
                yticklabels={}, % Hide y tick labels
                %ytick=\empty, % Hide y ticks completely
			xmin=1.0, xmax=3.5,
			ymin=1e-3, ymax=1.0,
			xtick={1.0,1.5,2,2.5,...,4.0},
			legend pos = south west,
			ymajorgrids=true,
			xmajorgrids=true,
			grid style=dashed,
			legend style={font=\fontsize{6}{7}\selectfont,fill opacity=0.8,text opacity=1,legend columns=2},
			]
%0 plot for NMS (FER)
\addplot[
color=magenta,
mark=x,
very thin
]
coordinates {
(1.0,0.60277)
(1.5,0.45918)
(2.0,0.30066)
(2.5,0.18054)
(3.0,0.08475)
(3.5,0.03495)
};	
\addlegendentry{NMS(4)}

%1 plot for {NMS(4)+UDE(-1.0)+OSD($p=2$)}
\addplot[
color=blue,
mark= square,
very thin
]
coordinates {
(1.000, 0.445)
(1.500, 0.308)
(2.000, 0.182)
(2.500, 0.099)
(3.000, 0.041)
(3.500, 0.016)
};	
\addlegendentry{N(4)-U(-1.0)-O(2)}
%2 plot for {NMS(4)+UDE(0.5)+OSD($p=2$)}
\addplot[
color=red,
mark=*,
very thin
]
coordinates {
(1.000, 0.390)
(1.500, 0.266)
(2.000, 0.145)
(2.500, 0.072)
(3.000, 0.026)
(3.500, 0.009)
};	
\addlegendentry{N(4)-U(0.5)-O(2)}
%0.5plot for ML (FER)
\addplot[
color=black,
mark=diamond,
very thin
]
coordinates {
(0.00, 6.329e-01)
(0.50, 4.975e-01)
(1.00, 3.704e-01)
(1.50, 2.445e-01)
(2.00, 1.447e-01)
(2.50, 7.353e-02)
(3.00, 2.595e-02)
(3.50, 7.918e-03)
(4.00, 2.134e-03)
(4.50, 4.751e-04)
(5.00, 5.337e-05)
(5.50, 6.300e-06)
};	
\addlegendentry{ML\cite{helmling19}}
%5 plot for {FER_FP-UDE(-1.)}
\addplot[
color=teal,
mark=triangle,
dashed,
very thin
]
coordinates {
(1.000, 0.19229)
(1.500, 0.14964)
(2.000, 0.09838)
(2.500, 0.06126)
(3.000, 0.02854)
(3.500, 0.01205)
};
\addlegendentry{FER$_{\text{UDE(-1.0)}}$}
%0.5plot for {FER_FP-UDE(0.5)}
\addplot[
color=purple,
mark=triangle*,
dashed,
very thin
]
coordinates {
(1.000, 0.04247)
(1.500, 0.03388)
(2.000, 0.02361)
(2.500, 0.01512)
(3.000, 0.00684)
(3.500, 0.00322)
};	
\addlegendentry{FER$_{\text{UDE(0.5)}}$}
\end{semilogyaxis}
\end{tikzpicture}
\caption{BCH (63,45) code}
\label{fer63_45_bch_ude}   
	\end{subfigure}
	\hfill
	\begin{subfigure}[t]{0.32\textwidth}
			\begin{tikzpicture}
		\begin{semilogyaxis}[
			width=10cm,
			height=7cm,
                scale = 0.6,
			xlabel={$E_b/N_0$(dB)},
             yticklabels={}, % Hide y tick labels
                %ytick=\empty, % Hide y ticks completely
			xmin=2.0, xmax=4.5,
			ymin=1e-3, ymax=1.0,
			xtick={2.0,2.5,3,3.5,...,4.5},
			legend pos = south west,
			ymajorgrids=true,
			xmajorgrids=true,
			grid style=dashed,
			legend style={font=\fontsize{6}{7}\selectfont,fill opacity = 0.7,text opacity = 1,legend columns=2},
			]
%0 plot for NMS (FER)
\addplot[
color=magenta,
mark=x,
very thin
]
coordinates {
(2.000, 0.53606)
(2.500, 0.40532)
(3.000, 0.26476)
(3.500, 0.15406)
(4.000, 0.07472)
(4.500, 0.03396)
};	
\addlegendentry{NMS(4)}

%1 plot for {NMS(4)+UDE(-1.0)+OSD($p=2$)}
\addplot[
color=blue,
mark= square,
very thin
]
coordinates {
(2.000, 0.511)
(2.500, 0.380)
(3.000, 0.248)
(3.500, 0.144)
(4.000, 0.069)
(4.500, 0.032)
};	
\addlegendentry{N(4)-U(-1.0)-O(2)}
%2 plot for {NMS(4)+UDE(0.5)+OSD($p=2$)}
\addplot[
color=red,
mark=*,
very thin
]
coordinates {
(2.000, 0.463)
(2.500, 0.331)
(3.000, 0.204)
(3.500, 0.113)
(4.000, 0.048)
(4.500, 0.022)
};	
\addlegendentry{N(4)-U(0.5)-O(2)}
%0.5plot for ML (FER)
\addplot[
color=black,
mark=diamond,
very thin
]
coordinates {
(0.00, 9.434e-01)
(1.00, 7.407e-01)
(2.00, 4.367e-01)
(3.00, 1.980e-01)
(4.00, 4.845e-02)
(5.00, 8.879e-03)
(5.50, 2.453e-03)
(6.00, 7.527e-04)
(6.50, 1.377e-04)
(7.00, 3.003e-05)
(7.50, 5.405e-06)
};	
\addlegendentry{ML\cite{helmling19}}
%5 plot for {FER_FP-UDE(-1.)}
\addplot[
color=teal,
mark=triangle,
dashed,
]
coordinates {
(2.000, 0.42362)
(2.500, 0.32605)
(3.000, 0.21799)
(3.500, 0.13215)
(4.000, 0.06540)
(4.500, 0.03087)
};
\addlegendentry{FER$_{\text{UDE(-1.0)}}$}
%0.5plot for {FER_FP-UDE(0.5)}
\addplot[
color=purple,
mark=triangle*,
dashed,
very thin
]
coordinates {
(2.000, 0.10885)
(2.500, 0.08442)
(3.000, 0.05831)
(3.500, 0.03676)
(4.000, 0.01901)
(4.500, 0.00963)
};	
\addlegendentry{FER$_{\text{UDE(0.5)}}$}
\end{semilogyaxis}
\end{tikzpicture}
\caption{RS (15,13) code}
\label{fer15_13_rs_ude}
	\end{subfigure}
	\caption{Impact on the FER of hybrid decoding schemes by the proposed UDE detection model for three short high-rate codes.}
	\label{fig:three_codes_ude}
\end{figure*}

We evaluate codes of different lengths from both LDPC and HDPC families. Specifically, we consider BCH codes with parameters (63,45), (127,64), and (127,99); bit-level high-rate RS codes (15,13) and (31,25); as well as LDPC Array (49,36) and CCSDS (128,64) codes. The corresponding PCMs for all codes are available in the channel codes database \cite{helmling19}. Open-source code for reproducing our results is publicly available on GitHub\footnote{\url{https://github.com/lgw-frank/Neural-Model-Augmented-Decoders}}, built on the TensorFlow platform. For a fair comparison, most reference decoding curves are taken directly from their original publications rather than re-implemented, to avoid potential discrepancies arising from unspecified implementation details or parameter settings.

For BCH and RS codes, the configurations of NMS decoding follow the descriptions in the previous section. For LDPC codes, $I_m=6$ and $I_m=10$ are applied to the LDPC (49,36) and (128,64) codes, respectively. Furthermore, we assume that the NMS, UDE, DIA, and SWA models are all well-trained, and that the decoding path initialized by \uc{ALMLT} for OSD is properly secured. At each tested SNR point (except for a few at high-SNR regime), adhering to community convention, at least 100 frame errors were collected to ensure statistical reliability.

\subsection{Decoding of Short High-Rate Codes}
In the hybrid decoding architecture, the NMS decoder triggers OSD only when it fails to converge to a codeword. However, UDEs in short high-rate codes often mislead the NMS decoder, producing false positives--valid codewords that differ from the transmitted ones. This motivates the proposed UDE detection model.

Fig.~\ref{fig:three_codes_ude} compares the FER performance with the UDE detection model disabled (denoted by UDE($-1.0$)) and enabled (denoted by UDE($0.5/0.9$)), where 'N(4)-U(-1.0)-O(2)' denotes the combination of NMS with $I_m=4$, the UDE detection model with $m_g=-1.0$, and order-2 OSD. The contribution of residual UDEs to the total FER is isolated and labeled as FER$_{\text{UDE(-1.0)}}$ and FER$_{\text{UDE(0.5/0.9)}}$ for the respective cases. Since the curves of FER$_{\text{UDE(0.5/0.9)}}$ lie consistently below that of FER$_{\text{UDE(-1.0)}}$, we infer that the UDE detection model plays a key role in discriminating NMS false positives--an effect that is more pronounced for BCH and RS codes than for LDPC codes. On the other hand, an order-2 conventional OSD following the NMS decoder closes the gap to ML performance in the presence of the UDE detection model, producing nearly overlapping FER curves, whereas some gap remains in the absence of the UDE detection model for BCH and RS codes.
\subsection{Decoding of Longer Codes}
Hereafter, we refer to the OSD method integrating the three distinct ingredients--namely, the DIA model, SWA model, and optimized decoding path--as DDS, and further denote DDS with parameter sets $(l_t,b_s,w_t,s_m)= (10^3,10^2,5,0.5)$ and $(10^4,10^2,5,0.5)$ as DDS1 and DDS2, respectively.

\begin{figure}[htbp]
	\centering
	\resizebox{\linewidth}{!}{%
		\begin{tikzpicture}
	\begin{semilogyaxis}[
		width=\linewidth,
		height=0.7\linewidth,
		scale only axis,
		xlabel={$E_b/N_0$ (dB)},
		ylabel={FER},
		xmin=.95, xmax=4.05,
		ymin=1e-5, ymax=1,
		xtick={1.0,1.5,2.0,2.5,3.0,3.5,4.0},
		legend pos=south west,
		ymajorgrids=true,
		xmajorgrids=true,
		grid style=dashed,
		legend style={font=\fontsize{7}{8}\selectfont, fill opacity=0.7, text opacity=1, legend columns=2},
		]
		
		% 1. NMS(10) - magenta, x marker
		\addplot[color=magenta, mark=x, thin,solid]
		coordinates {
			(1.00,0.83288) (1.25,0.76222) (1.50,0.67367) (1.75,0.58023)
			(2.00,0.47881) (2.25,0.37041) (2.50,0.27065) (2.75,0.18850)
			(3.00,0.11920) (3.25,0.07240) (3.50,0.04295) (3.75,0.02206) (4.00,0.01080)
		};	
		\addlegendentry{NMS(10)}
		
% TNMS(10) - brown, square, dashed
\addplot[color=brown, mark=square, thin, solid]
		coordinates {
			(1.00,0.77333) (1.50,0.63000) (2.00,0.44800) (2.50,0.23111)
			(3.00,0.10579) (3.50,0.03450) (4.00,0.00886) (4.50,0.00174)
		};	
		\addlegendentry{TNMS(10) \cite{ullah2011two}}
		
% TNMS(20) - olive, triangle, dotted  
\addplot[color=olive, mark=triangle, thin, solid]
		coordinates {
			(1.00,0.70667) (1.50,0.54500) (2.00,0.36333) (2.50,0.16154)
			(3.00,0.06182) (3.50,0.01950) (4.00,0.00346) (4.50,0.00056)
		};	
		\addlegendentry{TNMS(20) \cite{ullah2011two}}

		% 4. NBP(50) - red, o marker
		\addplot[color=red, mark=o, thin,solid]
		coordinates {
			(1.0,0.91) (1.5,0.71) (2.0,0.4) (2.5,0.2) (2.9,0.09)
			(3.0,0.07) (3.25,0.035) (3.65,0.01) (3.75,0.007) (4.0,0.003)
		};	
		\addlegendentry{NBP(50) \cite{buchberger21}}
		% 5. BP(40) - orange, + marker
\addplot[color=orange, mark=+,  thin,solid]
coordinates {
	(1.0,0.82) (1.5,0.6) (2.0,0.31) (2.5,0.15) (3.0,0.056) (3.5,0.013) (4.0,0.0021)
};	
\addlegendentry{BP(40) \cite{helmling19}}		
		% 6. NBP-D(10,4,4) - purple, halfcircle marker
		\addplot[color=purple, mark=halfcircle*,  thin]
		coordinates {
			(1.0,0.25) (1.5,0.12) (2.0,5e-2) (2.5,1.5e-2)
			(3.0,4e-3) (3.5,8e-4) (4.0,1.2e-4)
		};
		\addlegendentry{NBP-D(10,4,4) \cite{buchberger21}}
		
		% 7. NMS(10)-DDS1 - teal, triangle (solid)
		\addplot[color=teal, mark=square*, thin,solid]
		coordinates {
			(1.00,0.19526) (1.25,0.12843) (1.50,0.08322) (1.75,0.04763)
			(2.00,0.02750) (2.25,0.01379) (2.50,0.00633) (2.75,0.00310)
			(3.00,0.00125) (3.25,0.00050) (3.50,0.00016)
		};
		\addlegendentry{NMS(10)-DDS1}
		
		% 8. NMS(10)-DDS2 - cyan, triangle* (filled)
		\addplot[color=cyan, mark=triangle*,  thin,solid]
		coordinates {
			(1.00,0.13390) (1.25,0.07927) (1.50,0.05090) (1.75,0.02415)
			(2.00,0.01228) (2.25,0.00578) (2.50,0.00222) (2.75,0.00091)
			(3.00,0.00032) (3.25,0.00009)
		};
		\addlegendentry{NMS(10)-DDS2}
		
		% 9. MRB(4) - blue, * marker
		\addplot[color=blue, mark=*, thin]
		coordinates {
			(1.0,1.5e-1) (1.5,5e-2) (2.0,1.1e-2) (2.5,1.4e-3)
			(3.0,1.8e-4) (3.5,1.5e-5) (4.0,1.2e-6)
		};
		\addlegendentry{MRB(4) \cite{baldi2016use}}
		
		% 10. ML - black, diamond* marker
		\addplot[color=black, mark=diamond*,  thin]
		coordinates {
			(1.00,1.064e-01) (1.50,3.397e-02) (2.00,8.773e-03)
			(2.50,1.168e-03) (3.00,1.321e-04) (3.50,1.022e-05)
		};	
		\addlegendentry{ML \cite{helmling19}}
		
	\end{semilogyaxis}
\end{tikzpicture}%
	}
	\caption{FER of decoding schemes of LDPC (128,64) code}
	\label{ferber_ldpc_128_64_code}
\end{figure}

For the half-rate LDPC (128,64) code, the two-way NMS \cite{ullah2011two}, denoted as TNMS, assigns two trainable weights to the min-term of each check depending on the message sign alteration in consecutive iterations. 

The simulation results in Fig.~\ref{ferber_ldpc_128_64_code} show that the benchmarked NMS with $I_m=10$ (denoted as NMS(10)) falls slightly behind TNMS(10), and about 0.3 dB behind TNMS(20), whose FER curve nearly overlaps with that of NBP with $I_m=50$. While the latter slightly lags behind layered BP decoding with $I_m=40$, this suggests that the marginal FER improvement achieved by the neuralized adaptation of BP can be easily offset by deliberately scheduling the message update order of BP. NBP-D(10,4,4) \cite{buchberger21} takes a significant step toward the ML bound, achieving a 1.0 dB gain over NBP, albeit at the cost of much higher complexity and advanced decimation techniques that compromise the parallelizable data flow of BP. At a FER of $10^{-3}$, the hybrid NMS-DDS1 scheme outperforms NBP-D(10,4,4) by 0.5 dB. An additional 0.25 dB gain is achieved when upgrading the DDS component from DDS1 to DDS2, or equivalently, when the decoding path length $l_t$ is extended from $10^3$ to $10^4$. Within 0.1 dB of the ML curve \cite{helmling19}, MRB(4) \cite{baldi2016use}--essentially an order-4 OSD--outperforms the NMS-DDS2 scheme by only 0.15 dB in the high-SNR region, while overlapping with it in the low-SNR region below 2.0 dB.

However, compared with more than $6.7\times10^5$ TEPs in its decoding path for MRB(4), the $I_{at}$ of DDS2 drops sharply from approximately 4500 to 600 over the SNR range of 1.0-3.25 dB for all NMS decoding failures, despite a nominal $l_t=10^4$. For reference, the hybrid NMS with $I_m=10^2$ combined with a conventional OSD scanning an \uc{ALMLT} list \cite{baldi2016use}--omitted to avoid clutter in the figure--requires $l_t=2\times10^5$ to match MRB(4) in terms of FER performance. More importantly, in pursuing ML performance, much higher throughput is achieved for such NMS-OSD schemes due to the role of the NMS component decoder, which is well utilized in the proposed NMS-DDS scheme that simultaneously enjoys the evident reduction of $I_{at}$ in measuring OSD complexity. 

\begin{figure}[htbp]
	\centering
	\resizebox{\linewidth}{!}{%
		\centering
\begin{tikzpicture}
	\begin{semilogyaxis}[
		width=\linewidth,
		height=0.7\linewidth,
		scale only axis,
		xlabel={$E_b/N_0$ (dB)},
		ylabel={FER/BER},
		xmin=0.95, xmax=5.55,
		ymin=1e-5, ymax=1,
		xtick={1.0,1.5,2.0,2.5,3.0,3.5,4.0,4.5,5.0,5.5},
		legend pos=south west,
		ymajorgrids=true,
		xmajorgrids=true,
		grid style=dashed,
		legend style={font=\fontsize{7}{8}\selectfont, fill opacity=0.7, text opacity=1, legend columns=1},
		]
		
		% 1. BP-RNN(5) - violet, pentagon, dashed
		\addplot[
		color=violet,
		mark=pentagon,
		densely dashed,
		thick
		]
		coordinates {
			(2.00,1e-01)
			(3.00,7.5e-02)
			(4.00,4e-02)
			(5.00,1.3e-02)
			(6.00,3e-03)
		};	
		\addlegendentry{BP-RNN(5) \cite{nachmani18}}
		
		% 2. RNN-SS+PAN - teal, +, dashed
		\addplot[
		color=teal,
		mark=+,
		densely dashed,
		thick
		]
		coordinates {
			(2.0,0.1)
			(3.0,0.07)
			(4.0,0.04)
			(5.0,0.01)
			(6.0,8.9e-4)
		};
		\addlegendentry{RNN-SS+PAN \cite{lian2019learned}}
		
		% 3. NMS(8) for BER - blue, x, dashed 
		\addplot[
		color=blue,
		mark=x,
		densely dashed,
		thick
		]
		coordinates {
			(1.00,0.14971)
			(1.50,0.12646)
			(2.00,0.09713)
			(2.50,0.06509)
			(3.00,0.03576)
			(3.50,0.01519)
			(4.00,0.00489)
			(4.50,0.00104)
			(5.00,0.00014)
		};	
		\addlegendentry{NMS(8)}
		
		% 4. NMS(8) for FER - magenta, o marker (different from x), solid
		\addplot[
		color=magenta,
		mark=*,
		solid,
		thin
		]
		coordinates {
			(1.00,0.93495)
			(1.50,0.84881)
			(2.00,0.70094)
			(2.50,0.50150)
			(3.00,0.29223)
			(3.50,0.13072)
			(4.00,0.0428)
			(4.50,0.0098)
			(5.00,0.0014)
		};	
		\addlegendentry{NMS(8)}
		
		% 5. NMS(8)-DDS2 - red, square, solid
		\addplot[
		color=red,
		mark=square*,
		solid,
		thin
		]
		coordinates {
			(1.0,0.1333893)
			(1.5,0.0568703)
			(2.0,0.0165772)
			(2.5,0.0047392)
			(3.0,0.0008767)
			(3.5,7.06e-05)
		};	
		\addlegendentry{NMS(8)-DDS2}
		
		% 6. ISD-Dual(10^4) - orange, triangle, solid (halfcircle replaced)
		\addplot[
		color=orange,
		mark=triangle*,
		solid,
		thin
		]
		coordinates {
			(1.5,0.045)
			(1.75,0.025)
			(2.0,0.014)
			(2.25,6e-03)
			(2.5,2e-03)
			(2.75,8e-04)
			(3.0,3e-04)
			(3.25,8e-05)
			(3.5,2.5e-05)
		};	
		\addlegendentry{ISD-Dual($10^4$) \cite{bossert2022hard}}
		
		% 7. ML-LB - black, diamond, thick (highlighted as benchmark)
		\addplot[
		color=black,
		mark=diamond*,
		solid,
		thin
		]
		coordinates {
			(1.5,3.5e-02)
			(2.00,8e-03)
			(2.50,1.0e-03)
			(3.00,1.0e-04)
			(3.50,5.0e-06)
		};	
		\addlegendentry{ML-LB \cite{bossert2022hard}}
		
	\end{semilogyaxis}
\end{tikzpicture}%
	}
	\caption{FER of decoding schemes of BCH (127,64) code, where solid lines represent FER curves and dashed lines represent BER curves.}
	\label{ferber_127_64_bch_code}
\end{figure}

For the medium-rate BCH (127,64) code, shown in Fig.~\ref{ferber_127_64_bch_code}, achieving the ML lower bound (ML-LB) \cite{bossert2022hard} is more challenging. There exists about a 2.5 dB gap between NMS with $I_m=8$ and the ML-LB curves. ISD-Dual \cite{bossert2022hard} comes within 0.2 dB of the ML-LB by generating approximately $10^4$ TEPs on-the-fly via minimum-weight codewords in the dual space to update bit reliability measurements, at the cost of a throughput bottleneck due to Gaussian elimination and the repeated TEP generation process for OSD. In contrast, the proposed NMS-DDS2 scheme, with a fixed decoding path of nominal length $l_t=10^4$, lags behind with a gap of about 0.4 dB to the ML-LB. Regarding BER, RNN+SS+PAN \cite{lian2019learned} slightly leads BP-RNN \cite{nachmani18}--a neural BP variant--only in the high-SNR region, while both lag behind the proposed NMS by at least 1.3 dB at a FER of $10^{-2}$; this gap continues to widen with increasing SNR due to the steeper slope of the NMS curve.

\begin{figure}[htbp]
	\centering
	\resizebox{\linewidth}{!}{%
		\centering
\begin{tikzpicture}
	\begin{semilogyaxis}[
		width=\linewidth,
		height=0.7\linewidth,
		scale only axis,
		xlabel={$E_b/N_0$ (dB)},
		ylabel={FER/BER},
		xmin=0.95, xmax=5.05,
		ymin=1e-4, ymax=1,
		xtick={1.0,1.5,2.0,2.5,3.0,3.5,4.0,4.5,5.0,5.5},
		legend pos=south west,
		ymajorgrids=true,
		xmajorgrids=true,
		grid style=dashed,
		legend style={font=\fontsize{7}{8}\selectfont, fill opacity=0.7, text opacity=1, legend columns=1},
		]
		
		% 1. BP-RNN(5) (BER) - teal, +, dashed
		\addplot[
		color=teal,
		mark=+,
		densely dashed,
		thick
		]
		coordinates {
			(1.0,0.08)
			(2.0,0.06)
			(3.0,0.035)
			(4.0,0.016)
			(5.0,0.004)
			(6.0,5.5e-4)
		};
		\addlegendentry{BP-RNN(5) \cite{nachmani18}}
		
		% 2. NMS(5) (BER) - blue, x, solid (changed from dashed)
		\addplot[
		color=blue,
		mark=x,
		densely dashed,
		thick
		]
		coordinates {
			(1.00,0.0932)
			(1.50,0.0811)
			(2.00,0.0695)
			(2.50,0.0519)
			(3.00,0.0378)
			(3.50,0.0213)
			(4.00,0.0087)
			(4.50,0.00333)
			(4.75,0.00167)
			(5.00,0.00096)
			(5.25,0.00045)
		};	
		\addlegendentry{NMS(5)}
		
		% 3. NMS(8) (FER) - magenta, o (circle), solid
		\addplot[
		color=magenta,
		mark=*,
		solid,
		thin
		]
		coordinates {
			(1.00,0.97777)
			(1.50,0.93065)
			(2.00,0.82760)
			(2.50,0.68020)
			(3.00,0.47643)
			(3.50,0.27441)
			(4.00,0.11829)
			(4.50,0.04015)
			(4.75,0.02102)
			(5.00,0.00969)
			(5.25,0.00447)
		};	
		\addlegendentry{NMS(8)}
		
		% 4. NMS(8)-DDS1 - red, square, solid
		\addplot[
		color=red,
		mark=square*,
		solid,
		thin
		]
		coordinates {
			(1.0,0.684439)
			(1.5,0.4955711)
			(2.0,0.264832)
			(2.5,0.1124915)
			(3.0,0.0354654)
			(3.5,0.0076313)
			(4.0,0.000841)
			(4.5,0.0001)
		};	
		\addlegendentry{NMS(8)-DDS1}
		
		% 5. HOSD - violet, pentagon, dashed
		\addplot[
		color=violet,
		mark=pentagon*,
		solid,
		thin
		]
		coordinates {
			(2.00,3e-01)
			(2.50,1e-01)
			(3.00,3e-02)
			(3.50,6e-03)
			(4.00,8e-04)
			(4.50,7.5e-05)
		};	
		\addlegendentry{HOSD \cite{bailon2022concatenated}}
		
		% 6. OSD(2) - orange, triangle, solid (halfcircle replaced)
		\addplot[
		color=orange,
		mark=triangle*,
		solid,
		thin
		]
		coordinates {
			(1.0,0.48424)
			(1.5,0.33734)
			(2.0,0.19333)
			(2.5,0.084)
			(3.0,0.03017)
			(3.5,0.00637)
			(4.0,0.00086)
			(4.5,8.2e-05)
		};	
		\addlegendentry{OSD(2)}
		
		% 7. ML - black, diamond, thick (highlighted as benchmark)
		\addplot[
		color=black,
		mark=diamond*,
		solid,
		thin
		]
		coordinates {
			(2.00,2.551e-01)
			(2.50,1.020e-01)
			(3.00,2.680e-02)
			(3.50,6.907e-03)
			(4.00,8.074e-04)
			(4.50,4.958e-05)
			(5.00,2.984e-06)
		};	
		\addlegendentry{ML \cite{helmling19}}
		
	\end{semilogyaxis}
\end{tikzpicture}%
	}
	\caption{FER/BER of decoding schemes of BCH (127,99) code, where solid lines represent FER curves and dashed lines represent BER curves.}
	\label{ferber127_99_bch_code}
\end{figure}

For the high-rate BCH (127,99) code, the FER curves for NMS-DDS1, HOSD, and conventional OSD all approach the ML curve indistinguishably, as shown in Fig.~\ref{ferber127_99_bch_code}. This occurs because OSD--whether standalone or preceded by another decoder such as an algebraic decoder in HOSD \cite{bailon2022concatenated} or NMS in our proposal--will eventually reach the ML bound provided the number of traversed TEPs is sufficient to close the FER gap. Therefore, under the assumption that the preceding decoder is less complex than OSD, addressing throughput or complexity becomes more meaningful when comparing various hybrid schemes. The HDD component of HOSD boasts the simplest complexity but also suffers from the worst FER performance, which implies more frequent OSD calls to compensate performance to the ML bound, making it less appealing than the proposed NMS-DDS1 hybrid in scenarios requiring high throughput. In terms of BER, NMS(5) substantially outperforms BP-RNN(5), and this performance gap increases when the $I_m$ of both decoders is raised to 8.

\begin{figure}[htbp]
	\centering
	\resizebox{\linewidth}{!}{%
		\centering
\begin{tikzpicture}
	\begin{semilogyaxis}[
		width=\linewidth,
		height=0.7\linewidth,
		scale only axis,
		xlabel={$E_b/N_0$ (dB)},
		ylabel={FER},
		xmin=1.95, xmax=5.55,
		ymin=1e-5, ymax=1,
		xtick={2.0,2.5,3.0,3.5,4.0,4.5,5.0,5.5},
		legend pos=south west,
		ymajorgrids=true,
		xmajorgrids=true,
		grid style=dashed,
		legend style={font=\fontsize{7}{8}\selectfont, fill opacity=0.7, text opacity=1, legend columns=1},
		]
		
		% 1. HDD - magenta, x, solid
		\addplot[
		color=magenta,
		mark=x,
		solid,
		thin
		]
		coordinates {
			(2.5,0.95)
			(3.0,0.8)
			(3.5,0.61)
			(4.0,0.45)
			(4.5,0.25)
			(5.0,0.11)
			(5.5,0.045)
			(6.0,0.012)
		};
		\addlegendentry{HDD \cite{helmling19}}
		
		% 2. NMS(8) - violet, +, solid
		\addplot[
		color=violet,
		mark=+,
		solid,
		thin
		]
		coordinates {
			(2.00,0.89798) 
			(2.50,0.75432) 
			(3.00,0.54176) 
			(3.50,0.31298) 
			(4.00,0.13409) 
			(4.50,0.04243)
			(4.70,0.0248)
			(5.00,0.00903)
			(5.20,0.0046)
			(5.40,0.00215)
		};	
		\addlegendentry{NMS(8)}	
		
		% 3. HD-P-ABP - teal, square, dashed (distinct line style)
		\addplot[
		color=teal,
		mark=square*,
		solid,
		thin
		]
		coordinates {
			(4.0,3.2e-3)
			(4.5,3.35e-4)
			(5.0,2.35e-5)
		};	
		\addlegendentry{HD-P-ABP \cite{deng2020perturbed}}
		
		% 4. NMS(8)-DDS1 - blue, pentagon, solid
		\addplot[
		color=blue,
		mark=pentagon*,
		solid,
		thin
		]
		coordinates {
			(2.0,0.3951112)
			(2.5,0.1810368)
			(3.0,0.054436)
			(3.5,0.0120967)
			(4.0,0.0016708)
			(4.5,0.0001715)
		};	
		\addlegendentry{NMS(8)-DDS1}
		
		% 5. OSD(2) - red, triangle, dotted (distinct line style)
		\addplot[
		color=red,
		mark=triangle*,
		solid,
		thin
		]
		coordinates {
			(2.0,0.41)
			(2.5,0.18167)
			(3.0,0.05556)
			(3.5,0.01111)
			(4.0,0.00144)
			(4.5,1.3e-04)
		};	
		\addlegendentry{OSD(2)}
		
		% 6. ML - black, diamond, thick (highlighted as benchmark)
		\addplot[
		color=black,
		mark=diamond*,
		solid,
		thin
		]
		coordinates {
			(2.50,1.562e-01)
			(3.00,4.299e-02)
			(3.50,8.786e-03)
			(4.00,1.009e-03)
			(4.50,8.562e-05)
			(5.00,5.048e-06)
			(5.10,2.999e-06)
			(5.20,1.336e-06)
			(5.30,9.115e-07)
		};	
		\addlegendentry{ML \cite{helmling19}}
		
	\end{semilogyaxis}
\end{tikzpicture}%
	}
	\caption{FER of decoding schemes of RS (31,25) code}
	\label{fer_rs_31_25_code}
\end{figure}

For the high-rate RS (31,25) code, as shown in Fig.~\ref{fer_rs_31_25_code}, HDD, despite its low complexity, is significantly inferior to NMS with $I_m=8$ in terms of FER performance. The hybrid of this NMS with DDS1 performs nearly as well as order-2 OSD, which itself is within 0.1 dB of the ML curve. In comparison, HD-P-ABP \cite{deng2020perturbed} employs an advanced message scheduling strategy for BP based on received LLRs, sacrificing parallelizability, yet still shows minor performance degradation relative to the proposed hybrid scheme.

From the above simulation results for LDPC, RS, and BCH codes, one key finding is that NMS consistently exhibits a curve slope comparable to the various ML curves. This implies that the performance gap between NMS and ML will not widen with SNR. In contrast, the flatter curves of NBP variants reveal a fundamental limitation that would resort to frequent OSD post-processing when targeting the ML bound.

Concerning the LDPC (128,64) and BCH (127,64) codes of similar rates, a close examination of Figs.~\ref{ferber_ldpc_128_64_code} and \ref{ferber_127_64_bch_code} reveals that both codes demonstrate similar ML curves in the SNR region below 3.0 dB. Thus, the LDPC code is favored in this region because its hybrid NMS-OSD scheme can approach the ML bound at the cost of lower complexity, which is mostly attributed to its NMS component. However, in the high-SNR region, the BCH code is likely to take over the advantage due to its better FER performance, as can be inferred from the steeper ML curve of the former. Another distinct advantage of the proposed hybrid scheme is its ready extensibility to longer codes. In contrast, most dedicated decoding schemes in the literature are limited to codes with block lengths below 100 and are hardly scalable due to the curse of dimensionality known in the community.
\subsection{Complexity Analysis} 
Two metrics are vital to evaluate the average complexity of the proposed hybrid scheme: the average number of iterations $I_{an}$ required by the NMS decoder, and the average number of TEPs $I_{at}$ required by the SWA-based OSD.

\begin{figure}[htbp]
	\centering
	\resizebox{\linewidth}{!}{%
		\includegraphics{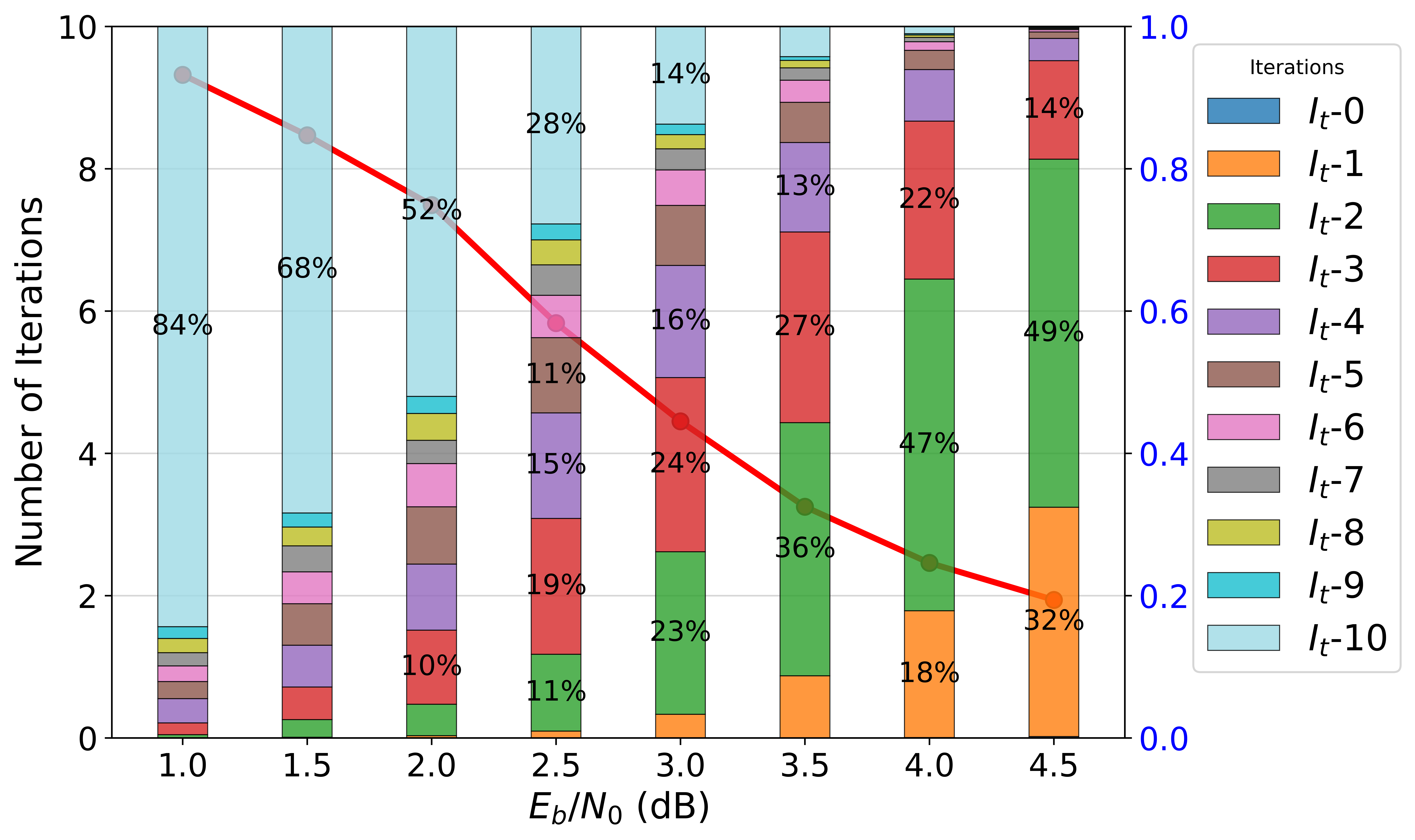}}%
	\caption{Distribution of number of iterations required for NMS ($I_m=10$) decoding convergence of LDPC (128,64) code.}
	\label{fig:iterations_ldpc_128_64}
\end{figure}

For the LDPC (128,64) code, given $I_m=10$ and early stopping enabled via \eqref{eq_early_termination}, Fig.~\ref{fig:iterations_ldpc_128_64} shows that the $I_{an}$ curve of NMS decoding decreases from 9.31 at SNR $=1.0$~dB to 1.94 at SNR $=4.5$~dB. The stacked bars illustrate the iteration distribution at each SNR point, with only fractions above 20\% explicitly annotated. At SNR $=1.0$~dB, 84\% of sequences require the full 10 iterations. As SNR increases, the distribution shifts mass toward lower iteration counts until, at SNR $=4.5$~dB, 95\% of sequences converge within 3 iterations.

\begin{figure}[htbp]
	\centering
	\resizebox{\linewidth}{!}{%
		\includegraphics{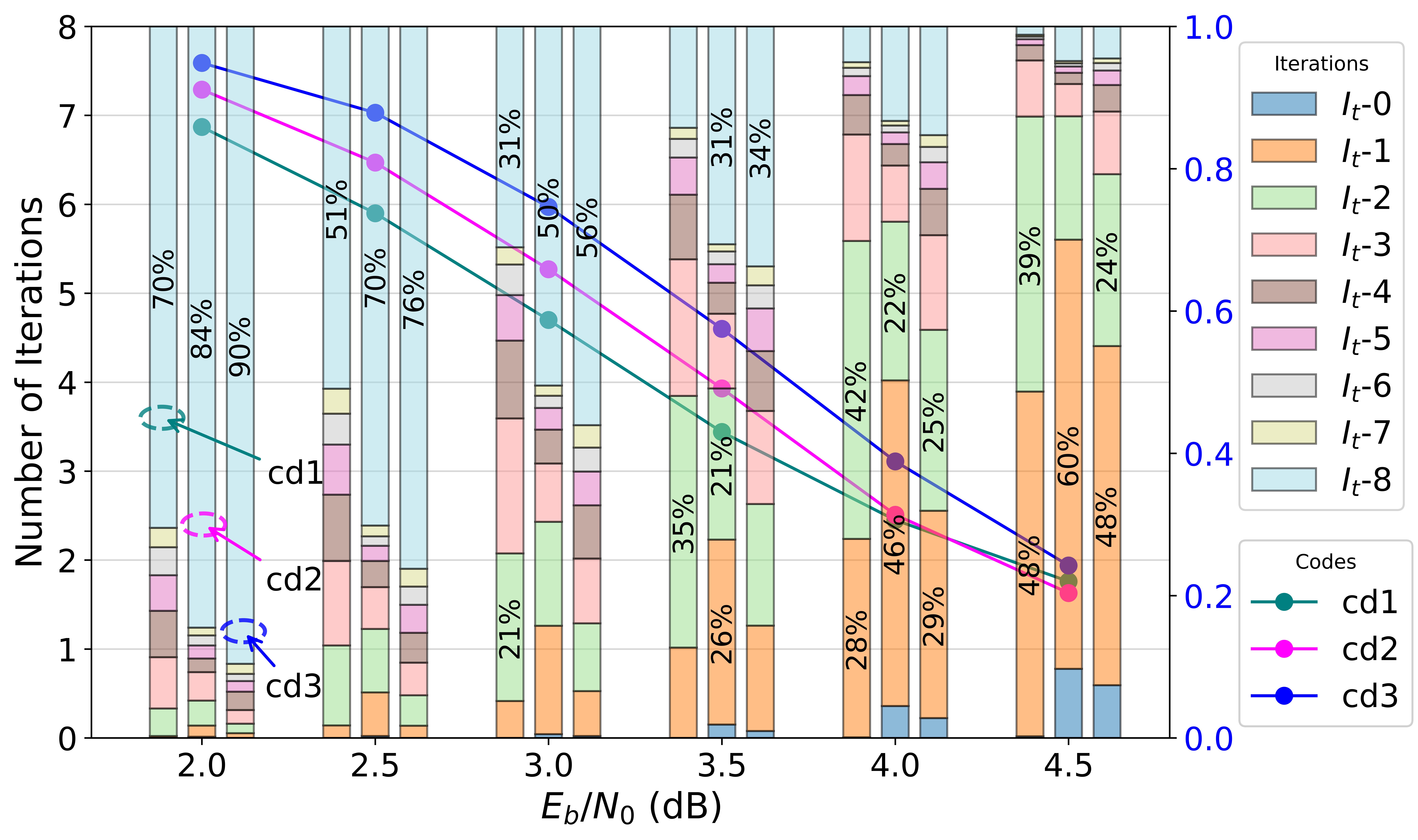}%
	}
	\caption{Distribution of number of iterations required for NMS ($I_m=8$) decoding convergence of BCH (127,64), (127,99) codes and RS (31,25) code.}
	\label{fig:iterations_hdpc_bch_rs}
\end{figure}

For HDPC codes, including BCH (127,64), BCH (127,99), and the binary image of RS (31,25) (denoted as cd1, cd2, and cd3, respectively), Fig.~\ref{fig:iterations_hdpc_bch_rs} demonstrates a similar trend for NMS ($I_m=8$) decoding convergence. As SNR increases from 2.0~dB to 4.5~dB, $I_{an}$ decreases from above 7 to nearly 2 for all codes, with cd2 showing the fastest convergence. The stacked bars again confirm that higher SNRs lead to the dominance of low-iteration decoding across all codes.

\begin{figure}[htbp]
	\centering
	\includegraphics[width=0.45\textwidth]{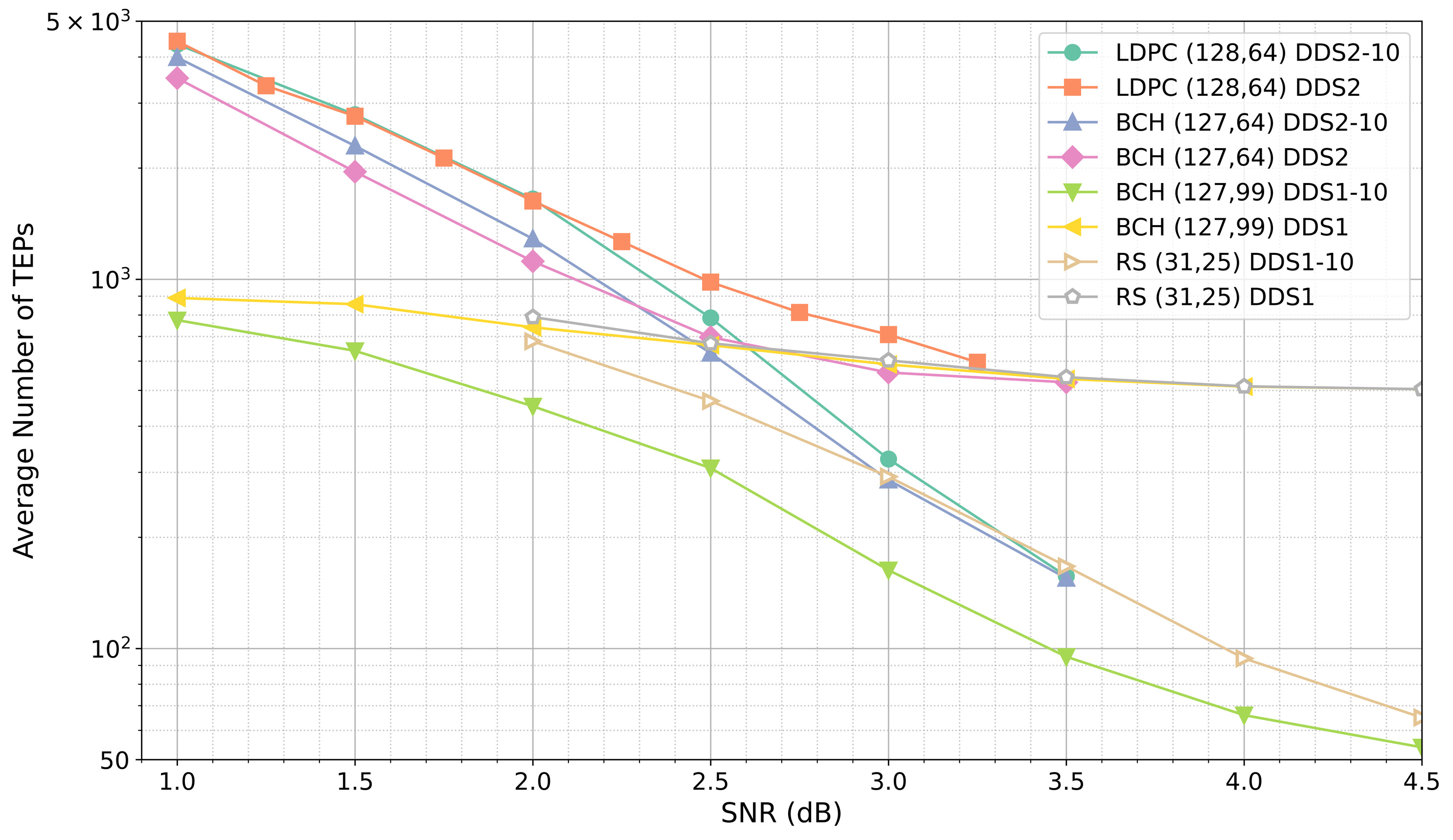}
	\caption{Average number of TEPs $I_{at}$ required for OSD decoding of NMS decoding failures for various codes.}
	\label{fig:iterations_osd_codes}
\end{figure}

Let the block size $b_s$ be reduced from the default 100 to 10 for the decoding paths of both DDS1 and DDS2, denoted as DDS1-10 and DDS2-10, respectively. As shown in Fig.~\ref{fig:iterations_osd_codes}, for decoding paths of length $l_t=10^4$, SWA-based early termination reduces $I_{at}$ to around 4000 starting at SNR $=1.0$~dB for DDS2 and DDS2-10 applied to LDPC (128,64) and BCH (127,64) codes. As SNR increases beyond 2.5~dB, the DDS2 curves saturate near the theoretical lower bound $b_s w_t = 500$, i.e., the first $w_t$ blocks of TEPs involved in preparing the initial window of the SWA model, where $w_t=5$ is assumed. In comparison, the lower bound $b_s w_t = 50$ is expected for DDS2-10. For high-rate codes BCH (127,99) and RS (31,25), DDS1 and DDS1-10 similarly reduce $I_{at}$ from a nominal $10^3$ to respective bounds of 500 and 50 in the high-SNR region. Notably, these gains are achieved without observable FER degradation, validating the SWA model's effectiveness in reducing computational complexity.

Omitting the FER attributed to UDEs for simplicity, for the NMS-OSD hybrid scheme, denote $F_n$ as the FER of NMS and $F_o$ as the FER of OSD when processing NMS decoding failures. The comprehensive FER of the hybrid scheme is $F_{hb} = F_n F_o$, while its BER is approximated as $\frac{1-R}{2} F_{hb}$, assuming half the length of the LRB contains erroneous bits and omitting the at-most $p$ bits in the MRB due to incorrectly decoded codewords by OSD.

At a given SNR, the average complexity $C_{hb}$ of a sequentially hybrid scheme is expressed as \cite{baldi2016use}:
\begin{equation}
	C_{hb} = C_{n} + F_{n} C_{o},
	\label{weighted_complexity}
\end{equation}
where $C_{n}$ and $C_o$ denote the complexity of NMS and OSD, respectively. Following \cite{baldi2016use}, the average number of binary operations (BOPS) for LDPC codes is estimated as
\begin{align*}
	C_{n} &= I_{ai} N \left[ q \left( 5d_v + 2R + 1 \right) + 2d_v - 1 + R \right], \\
	C_{o} &\approx q N \log_2 N + \left(\tfrac{K}{2}\right)^3 + I_{at} \left( q\tfrac{N-K}{2} + \tfrac{NK}{2} \right),
\end{align*}
where $q=6$ quantization bits are assumed, $d_v$ is the average column weight of $\mathbf{H}$, and $R$ is the code rate. The complexity of OSD based on the generator matrix $\mathbf{G}$ is assumed equal to that based on the PCM $\mathbf{H}$. Concerning the approximation of $C_n$ for BCH and RS codes, the complexity ratio $C_r$ defined in the previous section should be treated as the multiplier of $C_n$ associated with baseline NMS when accounting for the computational load imposed by additional factors such as $I_s$ and $I_r$. Complexity contributions from DIA, SWA, and potential UDE models are also included in the hybrid scheme, though the deliberate design pursuing simplicity of the NN structures justifies omitting them from complexity calculations.

\begin{figure}[htbp]
	\centering
	\includegraphics[width=0.45\textwidth]{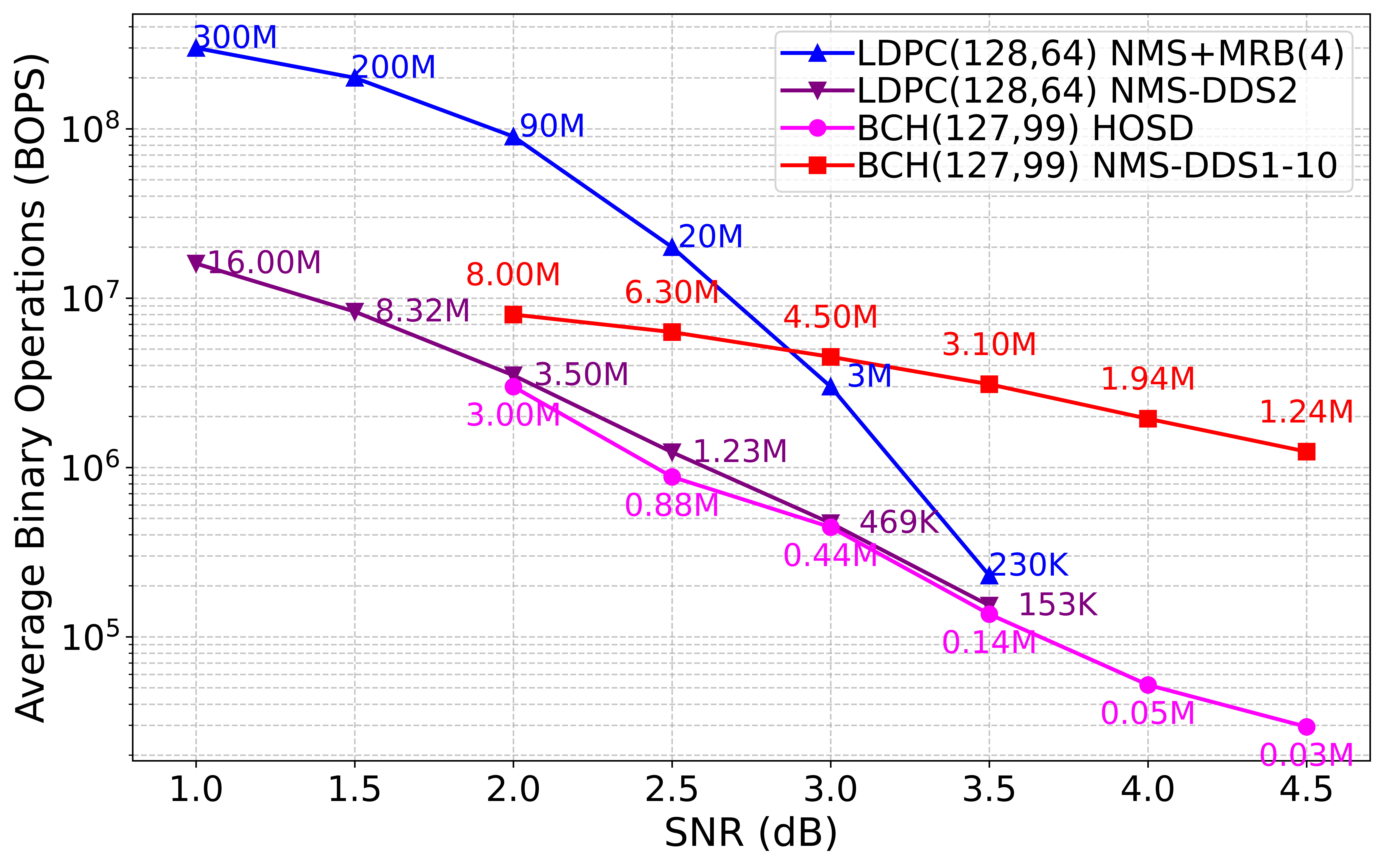}
	\caption{Average complexity of hybrid schemes across different codes in terms of BOPS.}
	\label{fig:hybrid_complexity}
\end{figure}

Fig.~\ref{fig:hybrid_complexity} compares the average complexity of various hybrid schemes under similar FER performance in terms of BOPS. For the LDPC (128,64) code, the proposed NMS-DDS2 requires significantly fewer BOPS than NMS+MRB(4) \cite{baldi2016use}. In the high-SNR region beyond 3.5 dB, both schemes converge to the dominant NMS complexity alone, since $F_n$ in \eqref{weighted_complexity} becomes negligible. For the BCH (127,99) code, the BOPS for HOSD--estimated from the reported weighted sum of FLOPS \cite{bailon2022concatenated} scaled by $q=6$--is much lower than our scheme in the high-SNR region, benefiting from the dominance of a simple HDD as the first-stage decoder.

Due to the bottlenecks of the mandatory Gaussian elimination operation and on-the-fly TEP ordering scheme, most OSD variants can hardly be accelerated on GPU/TPU. Our hybrid solution reduces its calls to as few as possible and uses a fixed TEP ordering for all received sequences. In terms of throughput, the proposed NMS-OSD hybrid benefits from the highly parallelizable and pipelined NMS component, as well as an optimized OSD. In principle, it will outperform ISD-Dual (a purely OSD variant) and other hybrids with weaker first-stage decoders such as HOSD across most SNR regions. Nevertheless, it remains to be validated on real GPU/TPU, FPGA, or ASIC hardware accelerators.  

\subsection{Optimization Framework}

The proposed hybrid decoding scheme introduces several parameters that affect both performance and complexity. In fact, every term in \eqref{weighted_complexity} is closely related to $\boldsymbol{\theta} = (I_s, I_r, I_m, l_t, s_m, w_t, b_s)$.

Given a target FER requirement $F^*$ at a specific SNR, the optimal parameter configuration can be formulated as the following constrained optimization problem:

\begin{equation}
	\label{eq:optimization_problem}
	\begin{aligned}
		\min_{\boldsymbol{\theta}} \quad C_{hb} \quad \text{s.t.} \quad F_{hb} \leq F^*,\quad \boldsymbol{\theta} \in \Theta,
	\end{aligned}
\end{equation}

where $\Theta$ denotes the feasible parameter space (e.g., $I_s, I_m, l_t, w_t, b_s \in \mathbb{Z}^+$, $I_r, s_m \in \mathbb{R}^+$). Due to the absence of closed-form expressions for $C_{hb}$ as a function of $\boldsymbol{\theta}$, the optimization problem in \eqref{eq:optimization_problem} cannot be solved analytically. A practical empirical approach would combine:
\begin{itemize}
	\item Grid search over the discrete variables $(I_s, I_m, l_t, w_t, b_s)$.
	\item Line search or Bayesian optimization over the continuous variables $I_r, s_m$.
\end{itemize}
Each evaluation requires Monte Carlo simulations with sufficient statistical reliability (e.g., at least 100 frame errors per SNR point). A full optimization campaign is a substantial undertaking beyond the scope of the current paper. Nevertheless, the formulation above provides a formal basis for reproducible trade-off tuning in future work or for practitioners seeking to adapt the proposed scheme to specific code and channel conditions.
\section{Conclusions and Future Research}
\label{conclusions}

\subsection{Conclusions}

Recent advances in BP-based decoding have shown that integrating diversified components can significantly improve FER performance. However, a major challenge lies in the severe throughput mismatch between decoding components when used in iterative mode. To address this, we favor a sequential architecture in which a high-throughput decoder processes the bulk of received sequences, forwarding only a small fraction of failure cases to a lower-throughput successor.

In our implementation, an efficient NMS decoder carries the majority of the decoding load due to its parallelizability, simplicity, and scalability to longer block lengths. For the small portion of failures left unresolved, the DIA model is first invoked to refine reliability measures from NMS trajectories. Then, OSD is applied to traverse TEPs along the periodically updated decoding path. Unless the decoding path is short enough for all TEPs to be evaluated in parallel, or minimal latency is the overriding concern, the SWA model assists OSD in making early termination decisions during traversal.

This hybrid architecture presents several distinctive characteristics: (1) the component decoders are naturally matched to data volume, with NMS efficiently handling large input streams while OSD focuses only on residual decoding failures; (2) each component decoder maximizes its strengths, as NMS prioritizes throughput and robust FER while OSD leverages optimized decoding paths and SWA-based early termination to minimize both computational and time complexity; and (3) for short high-rate codes, the UDE detection model further identifies false positives that pass parity checks in NMS decoding, enabling their post-processing in the second-stage OSD.

The collaboration among the decoding components yields decoder diversity gain and fulfills the design objective of balancing near-ML performance, high throughput, low latency, and moderate complexity. Furthermore, all decoding information is fully exploited: successful NMS outputs directly advance the decoding process, whereas its failure trajectories are fed into the DIA model to enhance OSD. For NMS, the decoding history supports the tuning of its normalization factor and the parameter learning for the UDE and DIA models; for OSD, historical records facilitate the optimization of the decoding path and the parameter learning for the SWA model.

\subsection{Limitations and Future Work}

As the current work is limited to software-level validation, critical hardware-aware analyses--including memory and compute costs, quantization effects, and hardware primitive mapping for NMS and OSD--are deferred to future investigation when hardware platforms become available.

Several directions warrant further exploration:

\begin{itemize}
	\item Improved UDE handling: For short high-rate block codes, most detected UDEs are still falsely decoded by OSD. Designing a new metric as an alternative to weighted distance to overcome this challenge is worth further investigation.
	
	\item Automorphism exploitation: For BCH and RS codes, discovering more automorphisms beyond the existing three types would be beneficial for lowering complexity by providing greater input diversity.
	
	\item Advanced neural architectures: Beyond the simple CNN-based DIA, more sophisticated neural architectures may provide more favorable bit-reliability estimates to unleash the full potential of OSD.
	
	\item Alternative first-stage decoders: To achieve a better balance between performance, throughput, and complexity, an alternative approach is to replace the NMS component in the hybrid scheme with TNMS \cite{ullah2011two}, which advocates flooding message scheduling and simple implementation.
	
	\item Theoretical framework: A rigorous theoretical framework linking TEP reduction to DIA ranking accuracy and error pattern statistics remains an open direction for further investigation.
	
	\item Robustness of neural-augmented decoders: The robustness of neural-augmented decoders under varying channel conditions and adversarial scenarios remains an open and important direction for our future research agenda.
\end{itemize}

\newpage
\bibliographystyle{IEEEtran}
\bibliography{main.bbl}
\end{document}